\documentclass[aps,prl,twocolumn,superscriptaddress,showpacs,preprintnumbers,amsmath,amssymb]{revtex4-2}
\usepackage{graphicx}% Include figure files
\usepackage{dcolumn}% Align table columns on decimal point
\usepackage{bm}% bold math
\usepackage{txfonts}
\usepackage{float}
\usepackage{latexsym}

\usepackage{color}
\usepackage[normalem]{ulem}
\begin{document}

%\preprint{APS/123-QED}

\title{Discriminating superconducting fluctuations from the pseudogap in Bi$_2$Sr$_2$Ca$_{n-1}$Cu$_n$O$_{2n+4+\delta} (n = 2,3)$:  A magnetotransport study} 
% In-plane normal state transport coefficients linked by a single parameter, the antiferromagnetic correlation length, in high-$T_c$ superconductors Bi$_2$Sr$_2$Ca$_{n-1}$Cu$_n$O$_{2n+4+\delta} (n = 2,3)$Force line breaks with \\Universal magnetotransport properties in hole-doped high-$T_c$ cuprates: BCS-BEC crossover scenario for the origin of the pseudogap

\author{Shunpei Yamaguchi}
\affiliation{Graduate School of Science and Technology, Hirosaki University, Hirosaki, Aomori, 036-8561 Japan}
\author{Nae Sasaki}
\affiliation{Graduate School of Science and Technology, Hirosaki University, Hirosaki, Aomori, 036-8561 Japan}
\author{Shintaro Adachi}
\affiliation{Graduate School of Science and Technology, Hirosaki University, Hirosaki, Aomori, 036-8561 Japan}
\affiliation{Faculty of Engineering/Graduate School of Engineering, Kyoto University of Advanced Science (KUAS), 615-8577 Japan}	
\author{Keiichi Harada}
\affiliation{Graduate School of Science and Technology, Hirosaki University, Hirosaki, Aomori, 036-8561 Japan}
\author{Yuki Teramoto}
\affiliation{Graduate School of Science and Technology, Hirosaki University, Hirosaki, Aomori, 036-8561 Japan}
\author{Shintaro Matsuda}
\affiliation{Graduate School of Science and Technology, Hirosaki University, Hirosaki, Aomori, 036-8561 Japan}
\author{Tomohiro Usui}
\affiliation{Graduate School of Science and Technology, Hirosaki University, Hirosaki, Aomori, 036-8561 Japan}

	%\affiliation{Graduate School of Science and Technology, Hirosaki University, 3 Bunkyo, Hirosaki, 036-8561 Japan}
	%\author{Mihaly M. Dobroka$^1$}
	%\affiliation{Graduate School of Science and Technology, Hirosaki University, 3 Bunkyo, Hirosaki, 036-8561 Japan}
%	\author{Shintaro Adachi}
% \affiliation{Nagamori Institute of Actuators, Kyoto University of Advanced Science (KUAS), Kyoto 615-8577, Japan}
        \author{Takenori Fujii}
\affiliation{Cryogenic Research Center, University of Tokyo, Bunkyo, Tokyo 113-0032, Japan}
\author{Takashi Noji}
\affiliation{Graduate School of Engineering, Tohoku University, Sendai 980-8579, Japan}			
	\author{Itsuhiro Kakeya}
\affiliation{Department of Electronic Science and Engineering, Kyoto University, Kyoto 615-8510, Japan}
\author{Haruka Taniguchi}
\affiliation{Graduate School of Engineering, Iwate University, Morioka 020-8551, Japan}
	%\affiliation{Department of Electronic Science and Engineering, Kyoto University, Kyoto 615-8510, Japan}
	\author{Michiaki Matsukawa}
\affiliation{Graduate School of Engineering, Iwate University, Morioka 020-8551, Japan}
	%\affiliation{Institute for Solid State Physics, University of Tokyo, 5-1-5 Kashiwanoha, Kashiwa, Chiba 277-8581, Japan}
	%\author{Koichi Kindo$^3$}
	%\affiliation{Institute for Solid State Physics, University of Tokyo, 5-1-5 Kashiwanoha, Kashiwa, Chiba 277-8581, Japan }
\author{Atsushi Miyake}
\affiliation{Institute for Solid State Physics, University of Tokyo, Kashiwa, Chiba 277-8581, Japan}
	\author{Hajime Ishikawa}
\affiliation{Institute for Solid State Physics, University of Tokyo, Kashiwa, Chiba 277-8581, Japan}
        \author{Koichi Kindo}
\affiliation{Institute for Solid State Physics, University of Tokyo, Kashiwa, Chiba 277-8581, Japan}
        %\author{Toshimitsu Ito}
%\affiliation{Research Institute for Advanced Electronics and Photonics, National Institute of Advanced Industrial Science and Technology (AIST), Higashi 1-1-1, Tsukuba, Ibaraki 305-8565, Japan}
\author{Takao Watanabe}
\email{E-mail address: watanabe.takao@nihon-u.ac.jp}
%\email{Present address: Physics Department, College of Engineering, Nihon University, Fukushima 963-8642, Japan. E-mail address: 
%watanabe.takao@nihon-u.ac.jp}
\affiliation{Graduate School of Science and Technology, Hirosaki University, Hirosaki, Aomori, 036-8561 Japan}
\affiliation{Physics Department, College of Engineering, Nihon University, Fukushima 963-8642, Japan}
\affiliation{Department of Advanced Materials Science, University of Tokyo, Kashiwa, Chiba 277-8561, Japan}
%        \author{Shojiro Kimura}
% \affiliation{Institute for Materials Research, Tohoku University, 2-1-1 Katahira, Aoba-ku, Sendai, 980-8577 Japan}
%	\author{Ken Hayama}
%\affiliation{Department of Electronic Science and Engineering, Kyoto University, Kyoto 615-8510, Japan}

	%\affiliation{Institute for Solid State Physics, University of Tokyo, 5-1-5 Kashiwanoha, Kashiwa, Chiba 277-8581, Japan}
	%\author{Koichi Kindo$^3$}
	%\affiliation{Institute for Solid State Physics, University of Tokyo, 5-1-5 Kashiwanoha, Kashiwa, Chiba 277-8581, Japan }
	%\author{Hajime Ishikawa}
%\affiliation{Institute for Solid State Physics, University of Tokyo, Kashiwa, Chiba 277-8581, Japan}
        %\author{Koichi Kindo}
%\affiliation{Institute for Solid State Physics, University of Tokyo, Kashiwa, Chiba 277-8581, Japan}
       % \author{Daniel S. Dessau}
% \affiliation{Department of Physics, University of Colorado at Boulder, Boulder, CO 80309, USA}

	%\affiliation{Institute for Materials Research, Tohoku University, 2-1-1 Katahira, Aoba-ku, Sendai, 980-8577 Japan}
%	\author{Takao Watanabe}
%\email{twatana@hirosaki-u.ac.jp}
%\affiliation{Graduate School of Science and Technology, Hirosaki University, Hirosaki, Aomori, 036-8561 Japan}
	%\affiliation{Graduate School of Science and Technology, Hirosaki University, 3 Bunkyo, Hirosaki, 036-8561 Japan}
%$^\star$\thanks{email: twatana@hirosaki-u.ac.jp}

	%\affiliation{Institute for Solid State Physics, University of Tokyo, 5-1-5 Kashiwanoha, Kashiwa, Chiba 277-8581, Japan$^3$}

%\author{Aaa Bee}

%\affiliation{Hirosaki University, Japan}


\date{\today}% It is always \today, today,

\begin{abstract}
Understanding the normal state is essential for uncovering the mechanism of high‑$T_c$ superconductivity. We investigate magnetotransport in Bi$_2$Sr$_2$CaCu$_2$O$_{8+\delta}$ and Bi$_2$Sr$_2$Ca$_2$Cu$_3$O$_{10+\delta}$ single crystals over a wide doping range. While the in‑plane resistivity and Hall coefficient show strong pseudogap‑induced temperature dependence, the $T^2$ Hall-angle behavior and the modified Kohler’s rule remain robust across all dopings. The onset temperatures of the pseudogap are clearly distinct from superconducting fluctuations, although they scale with the pseudogap magnitudes with a factor consistent with a $d$-wave superconductor. These results demonstrate that the pseudogap does not arise from superconducting fluctuations and instead suggest that it may originate from preformed Cooper pairing in the BCS–BEC crossover regime.

\end{abstract}

%\pacs{71.27.+a, 79.60.-i}% PACS, the Physics and Astronomy
\maketitle

%\section{INTRODUCTION}%%%INTRODUCTION%%%
The central challenge in copper‑oxide high-$T_c$ superconductors is the anomalous normal state, particularly the origin of the pseudogap \cite{Keimer15}. The pseudogap is an energy gap that opens above $T_c$, most prominently in underdoped samples. Its origin remains intensely debated, with proposals ranging from precursor superconductivity \cite{Li10,Wang06,Kaiser14}, to competing orders \cite{Tranquada95,Ghiringhelli12,Daou10}. Despite extensive experimental and theoretical efforts, its nature is still unresolved \cite{Timusk99,Hufner08,Kordyuk15,Vishik18}. 

Transport measurements in a magnetic field provide a powerful probe of the anomalous normal state. It is well established, near optimal doping, that the in‑plane resistivity $\rho_{ab}$ varies linearly with temperature, while the Hall coefficient $R_H$ scales approximately as $T^{-1}$, yielding a Hall angle $\cot \theta_H$ that follows a $T^2$ dependence \cite{Ong91}. This behavior was originally interpreted within the resonating-valence-bond framework as evidence for two distinct scattering times \cite{Ong91,Anderson91}, although this interpretation remains under debate.

Kontani later developed a Fermi‑liquid framework that incorporates strong antiferromagnetic spin fluctuations through current‑vertex-corrections (CVC)—the backflow term in Fermi‑liquid theory—within the fluctuation‑exchange (FLEX) approximation, hereafter referred to as the CVC theory \cite{Kontani08}. In the self‑consistent renormalization approach \cite{Moriya00}, the in-plane resistivity is given by
\begin{equation}
\rho_{ab} \propto \xi_{AF}^2 T^2, 
\end{equation}
where $\xi_{AF}$ is the antiferromagnetic correlation length. Within the CVC framework, the Hall coefficient satisfies \cite{Kontani08},
\begin{equation}
R_H \propto \xi_{AF}^2. 
\end{equation}
Since $\xi_{AF}^2 \propto T^{-1}$ in two-dimensional systems \cite{Moriya000}, the observed behaviors $\rho_{ab} \propto T$ and $R_H \propto T^{-1}$ near optimal doping follow naturally \cite{wata3}. Consequently, the Hall angle obeys
\begin{equation}
\cot \theta_H = \rho_{ab} / R_HB \propto T^2. 
\end{equation}
For the magnetoresistance $MR$, the modified Kohler's rule,
\begin{equation}
MR \propto \xi_{AF}^4B^2/\rho_{ab}^2 \propto \tan^2 \theta_H,
\end{equation}
is also derived \cite{Kontani08}. The CVC theory is consistent with expeimental results not only on hole-doped La$_{2-x}$Sr$_{x}$CuO$_4$ (LSCO) \cite{Ong95,Kimura96,Malinowski02} and YBa$_2$Cu$_3$O$_{7-\delta}$ (YBCO) \cite{Ong91,Ong95}, each at near optimal doping, but also on electron-doped La$_{2-x}$Ce$_{x}$CuO$_4$ (LCCO) \cite{Hussey25}. On the other hand, recent high-field studies on Bi$_2$Sr$_2$CuO$_{6+\delta}$ (Bi-2201), Tl$_{2}$Ba$_{2}$CuO$_{6+\delta}$ (Tl-2201), and LSCO demonstrate a new scaling for $MR$  \cite{Hussey21,Hussey24}, implying that the Fermi-liquid approach is insufficient for explaining the transport properties in hole-doped cuprates.

However, the applicability of the CVC theory to the underdoped regime of hole-doped systems remains uncertain, as many physical quantities exhibit anomalous behavior below $\approx$ 200 K due to the emergence of the strong pseudogap \cite{Kontani08}. Kontani examined the FLEX + $T$-matrix formalism—which incorporates scattering from superconducting fluctuations (SCF)—together with CVC, and showed that the anomalous transport coefficients in this temperature range can be reproduced \cite{Kontani02}. This result suggests that the pseudogap originates from SCF. In contrast, studies on HgBa$_{2}$CuO$_{4+\delta}$ (Hg-1201) have revealed that, within the pseudogap regime, the in-plane resistivity follows $\rho_{ab} \propto T^2$ \cite{Barisic13} and the magnetoresistance obeys the conventional Kohler’s rule ($MR \propto B^2/\rho_{ab}^2$) \cite{Barisic14}. These observations indicate that this region behaves as a conventional Fermi liquid, implying that the pseudogap is independent of SCF. To resolve this apparent inconsistency, it is therefore essential to investigate other systems containing CuO$_2$ planes with flatness comparable to that of Hg-1201.

\begin{figure}[t]
		\begin{center}
			\includegraphics[width=85mm]{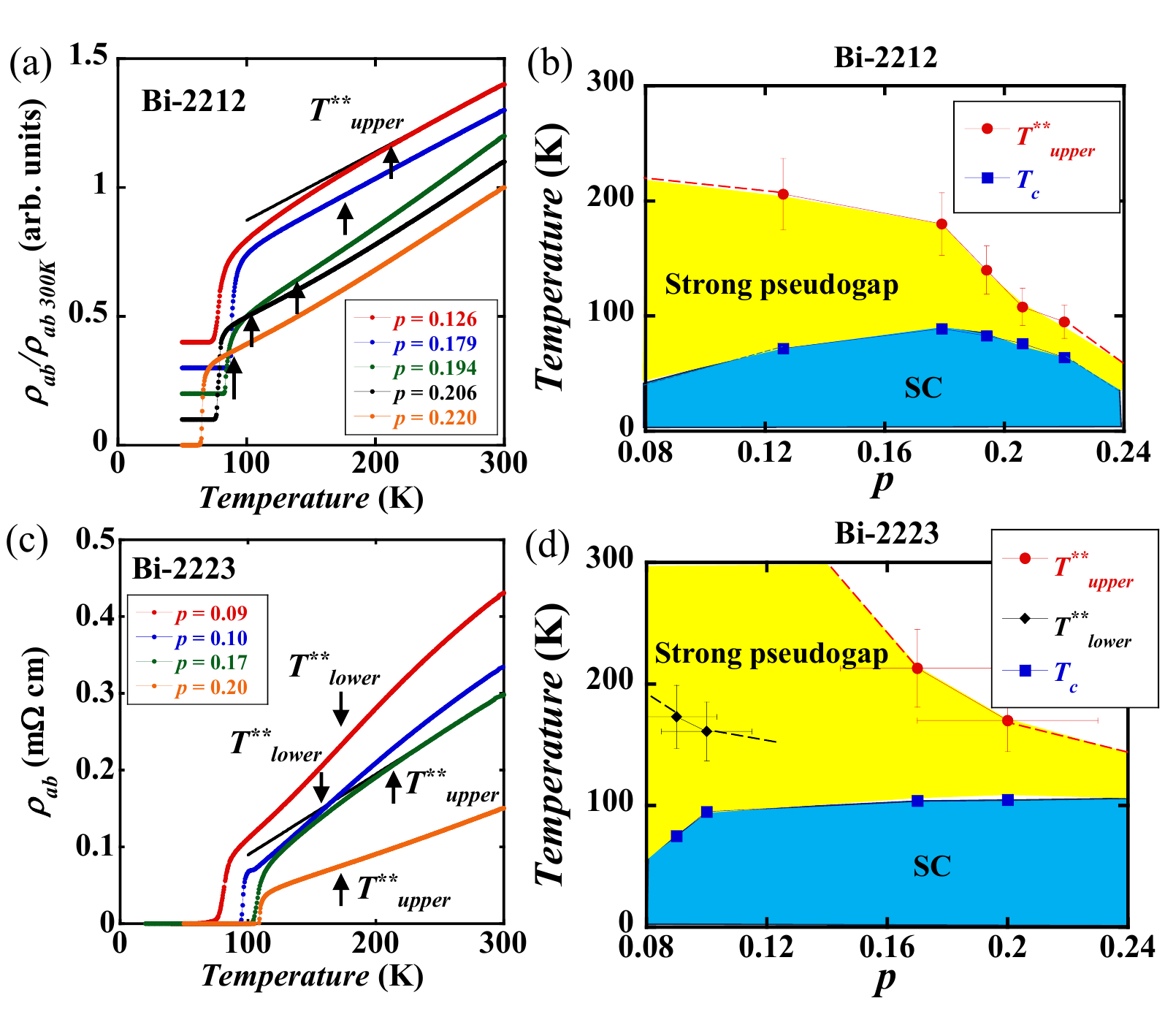}
			
			\caption{\label{fig1}(Color online) (a) Normalized in-plane resistivity $\rho_{ab} (T)$ for Bi-2212 single crystals with various doping levels. The curves are vertically shifted  for clarity. (b)  Doping dependence of $T^{**}_{upper}$ and $T_c$ for Bi-2212. (c) In-plane resistivity $\rho_{ab} (T)$ for Bi-2223 single crystals with various doping levels. (d) Doping dependence of $T^{**}_{upper}$, $T^{**}_{lower}$ and $T_c$ for Bi-2223. The arrows in (a) and (c) indicate the pseudogap temperatures $T^{**}_{upper}$ or $T^{**}_{lower}$. The solid straight lines in (a) and (c) represent linear extrapolations of the high-temperature $\rho_{ab} (T)$ and are included as guides to the eye. } 

		\end{center}
	\end{figure}

\begin{figure}[t]
		\begin{center}
			
			\includegraphics[width=85mm]{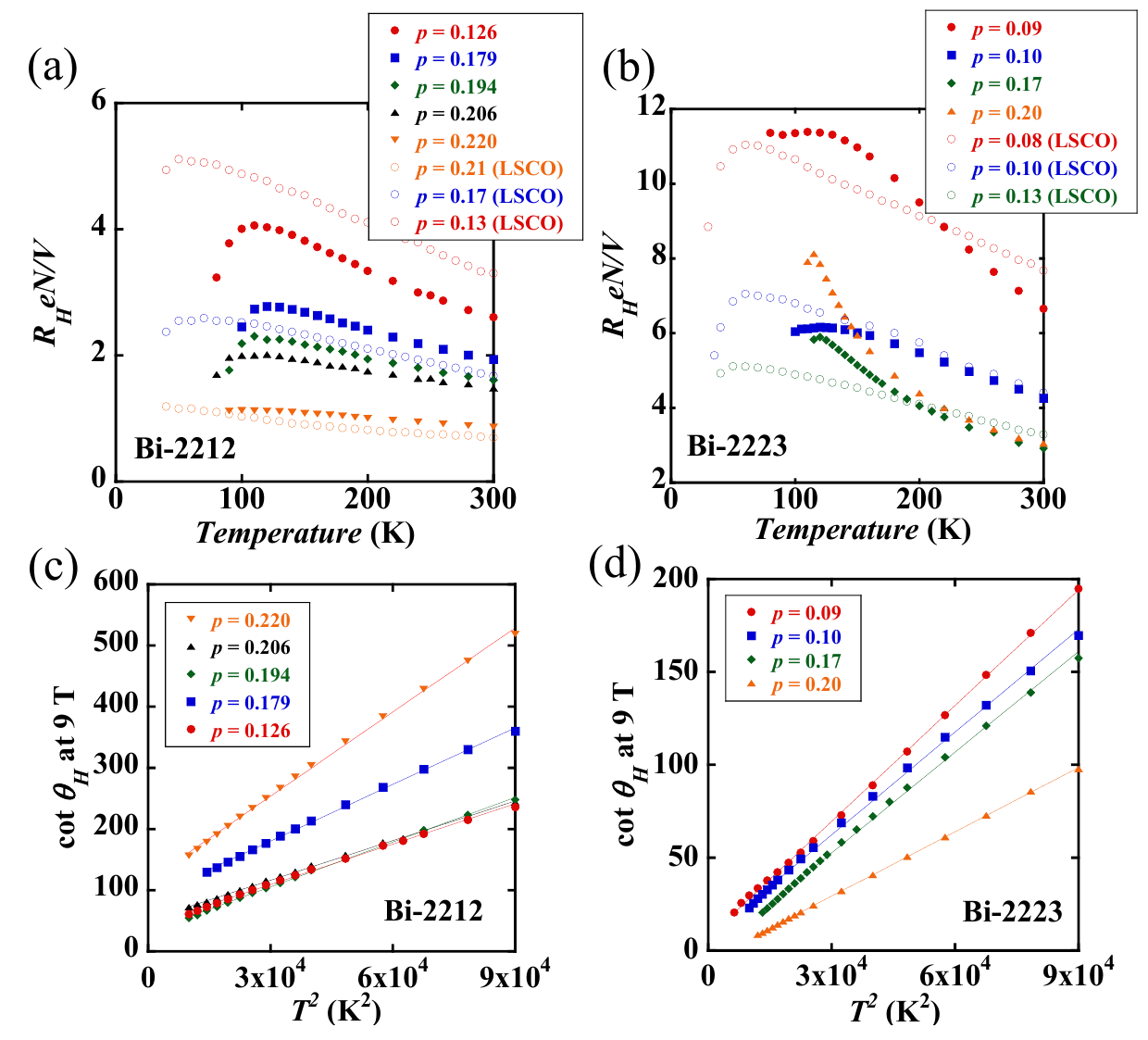}
			\caption{\label{fig2}(Color online) Normalized Hall coefficient $R_H (T)$ for (a) Bi-2212 and (b) Bi-2223 single crystals with various doping levels.  Data for LSCO are taken from Ref. \cite{Ando04}. Hall angle $\cot \theta_H$ at 9 T plotted against $T^2$ for (c) Bi-2212 and (d) Bi-2223 single crystals with various doping levels. }
		\end{center}
	\end{figure}

\begin{figure*}[t]
		\begin{center}
			
			\includegraphics[width=180mm]{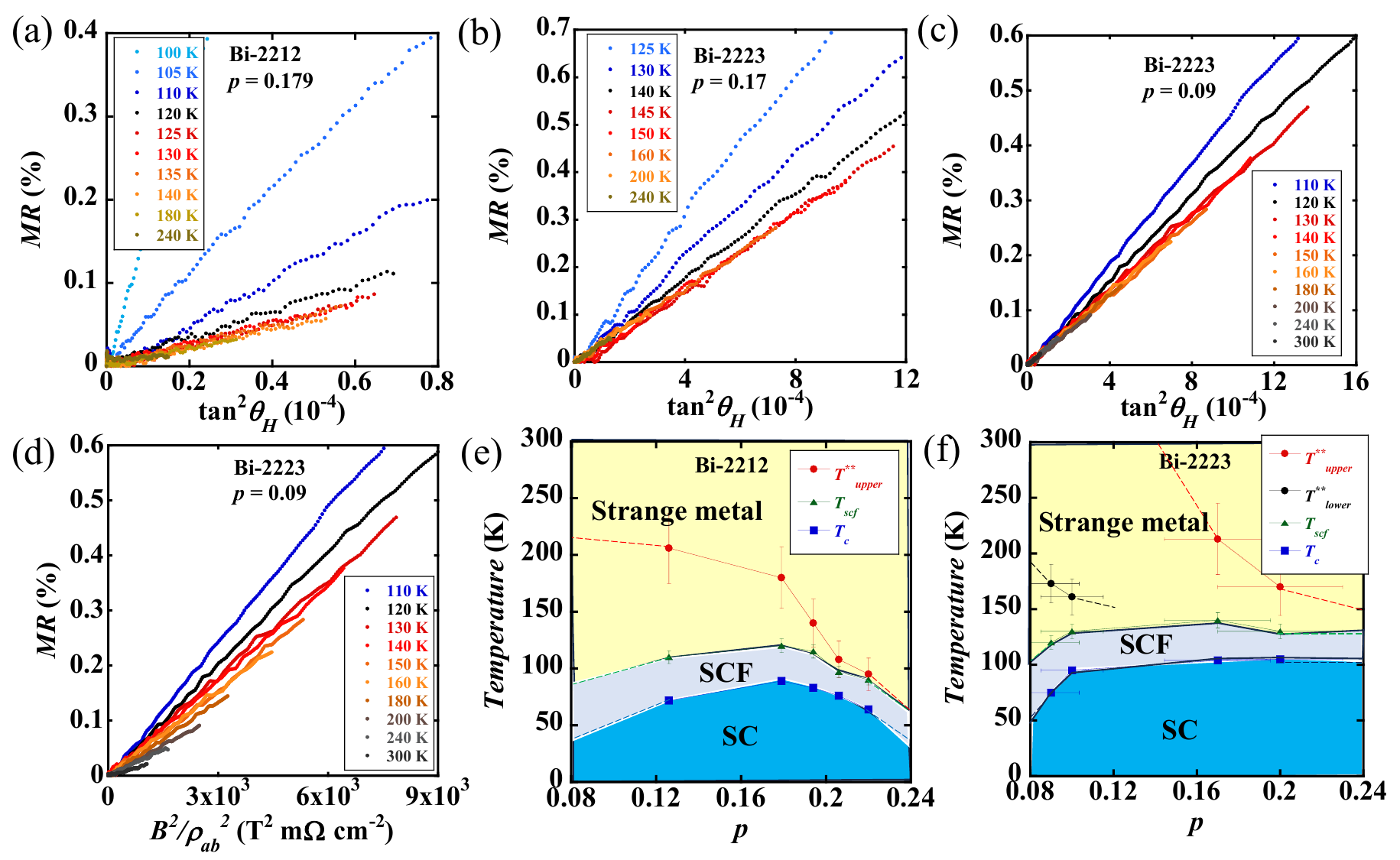}
			\caption{\label{fig3}(Color online) Magnetoresistance $MR$ versus $\tan^2 \theta_H$ at several temperatures for (a) Bi-2212 with $p$ = 0.179, (b) Bi-2223 with $p$ = 0.17, and (c) Bi-2223 with $p$ = 0.09. (d) $MR$ plotted against $B^2/\rho_{ab}^2$ for Bi-2223 with $p$ = 0.09. The temperatures corresponding to the black dots in panels (a)–(d) indicate $T_{scf}$. Phase diagrams for (e) Bi-2212 and (f) Bi-2223.}
		\end{center}
	\end{figure*}

Bi$_2$Sr$_2$Ca$_2$Cu$_3$O$_{10+\delta}$ (Bi-2223) provides an excellent platform for addressing this issue, as it contains an ideally flat inner CuO$_2$ plane and exhibits an optimal $T_c$ exceeding 100 K \cite{Fujii02}. To examine the material dependence, we also investigate Pb‑doped Bi$_2$Sr$_2$CaCu$_2$O$_{8+\delta}$ (Bi-2212) \cite{Watanabe22}. By comparing these two systems, we aim to clarify the role of the flat inner CuO$_2$ plane in generating the strong pseudogap .

For details of sample preparation and the determination of the doping level $p$,  see Supplemental Material \cite{Supplemental}. In‑plane magnetotransport measurements were performed using a Physical Property Measurement System (PPMS, Quantum Design) equipped with a Cernox thermometer. A five‑terminal configuration was used under magnetic fields up to 14 T ($B \parallel c$). Because the longitudinal magnetoresistance in high-$T_c$ cuprates is known to be small (below 10 \%) compared with the transverse component \cite{Kimura96,Heine99,Watanabe96}, we assume that the orbital contribution dominates the transverse $MR$. 

Figure 1(a) shows the temperture dependence of the normalized in-plane resistivity $\rho_{ab}(T)$ for Bi‑2212 single crystals with various doping levels. In the underdoped sample ($p$ = 0.126), a clear downward deviation from the high-temperature $T$-linear behavior is observed, reflecting a reduction in the scattering rate associated with the onset of the pseudogap \cite{Ito93}. We define the pseudogap temperature $T^{**}_{upper}$ as the point where $\rho_{ab}(T)$ falls by 1 $\%$ from the extrapolated high‑temperature linear behavior \cite{Watanabe97,Usui14} (see Supplemental Material \cite{Supplemental}). For the nearly optimally doped ($p$ = 0.179) and overdoped samples ($p$ = 0.194, 0.206, and 0.220), the high‑temperature resistivity is no longer $T$-linear. In these cases, $T^{**}_{upper}$ is defined as the temperature corresponding to the minimum in $d\rho_{ab}(T)/dT$ \cite{Watanabe22} (for plots of $d\rho_{ab}/dT$ versus $T$, together with complementary measurements that support our definition, see Supplemental Material \cite{Supplemental}). This $T^{**}_{upper}$ corresponds to the strong pseudogap temperature $T^{**}_{\rho_{ab}}$ defined in Ref. \cite{Watanabe22}. The resulting $T^{**}_{upper}$ values are plotted as a function of $p$ in Fig. 1(b). Notably, the pseudogap region persists well into the overdoped regime \cite{Watanabe22}.

Figure 1(c) shows $\rho_{ab} (T)$ for Bi-2223 at various doping levels. For the optimally doped ($p$ = 0.17) and overdoped ($p$ = 0.20) samples, $T^{**}_{upper}$ is defined in the same manner as for the corresponding Bi‑2212 samples. The underdoped samples ($p$ = 0.09 and 0.10) exhibit a $T^2$-like temperature dependence (for a plot of $\rho_{ab} (T)$ versus $T^2$, see Supplemental Material \cite{Supplemental}), similar to that reported for Hg‑1201 \cite{Barisic13}, with $T$-linear behavior appearing only above room temperature. To characterize the pseudogap in these samples, we take the temperature derivative of $\rho_{ab} (T)$ and define the temperature at which it reaches a maximum as $T^{**}_{lower}$ (for plots of $d\rho_{ab}/dT$ versus $T$, see Supplemental Material \cite{Supplemental}). This $T^{**}_{lower}$ corresponds to the temperature at which the pseudogap becomes sufficiently developed, matching $T_{pg}$ in Ref. \cite{Ando041} ($T^{**}_{lower}$ is equivalent to $T_{pg}$) and $T^{**}$ in Ref. \cite{Barisic13}. The resulting pseudogap temperatures are plotted as a function of $p$ in Fig. 1(d). Notably, the pseudogap region in Bi‑2223 is broader than that in Bi‑2212, consistent with the respective pseudogap magnitudes observed by angle-resolved photoemission spectroscopy (ARPES) \cite{Sato02}, interlayer tunneling spectroscopy \cite{Suzuki03,Suzuki12}, and optical spectroscopy \cite{Tajima24}. 

Figure 2(a) shows the temperature dependence of $R_H (T)$ for Bi-2212 at various doping levels. To facilitate comparison with LSCO \cite{Ando04}, the Hall coefficient is normalized as $R_HeN/V$, where $e$ is the electronic charge, $N$ is the number of Cu atoms per unit cell, and  $V$ is the unit-cell volume \cite{Ando04,Ando00}. Both the magnitude and temperature dependence closely resemble those of LSCO. In the underdoped sample, $R_HeN/V$ exhibits an approximate $T^{-1}$ increase upon cooling above $\approx$ 200 K; however, this trend is suppressed at lower temperatures and ultimately reverses. This behavior reflects the suppression of antiferromagnetic fluctuations due to the opening of the pseudogap, which limits the growth of the antiferromagnetic correlation length $\xi_{AF}$ \cite{Kontani08}.

Figure 2(b) shows the temperature dependence of $R_HeN/V$ for Bi-2223 at various doping levels. The magnitude decreases with increasing doping, similar to the trend in LSCO \cite{Ando04}. However, even at optimal doping, its value remains comparable to that of LSCO at $x$ = 0.13. In the overdoped samples, $R_HeN/V$ does not decrease further; instead, it increases relative to the optimally doped sample, despite clear evidence from the resistivity that doping continues to increase [Fig.1(c)]. This unusual behavior can be attributed to magnetic interactions between the inner CuO$_2$ plane (IP) and the outer CuO$_2$ planes (OP), which enhance the antiferromagnetic correlation length $\xi_{AF}$. These interlayer magnetic interactions are particularly strong in the overdoped regime, leading to the observed enhancement of $R_HeN/V$.

Figures 2(c) and 2(d) show the Hall angle for Bi‑2212 and Bi‑2223, respectively. Despite the strong pseudogap effects on both $\rho_{ab} (T)$ and $R_H (T)$ [Figs.1(a), 1(c), 2(a), and 2(b)], the Hall angle robustly follows $\cot \theta_H \propto T^2$ across $T^{**}_{upper}$ \cite{Barisic19}. This robustness suggests that the pseudogap is a normal-state property, however, this observation alone does not prove that the pseudogap is purely a normal‑state phenomenon.

\begin{figure}[t]
		\begin{center}
			
			\includegraphics[width=80mm]{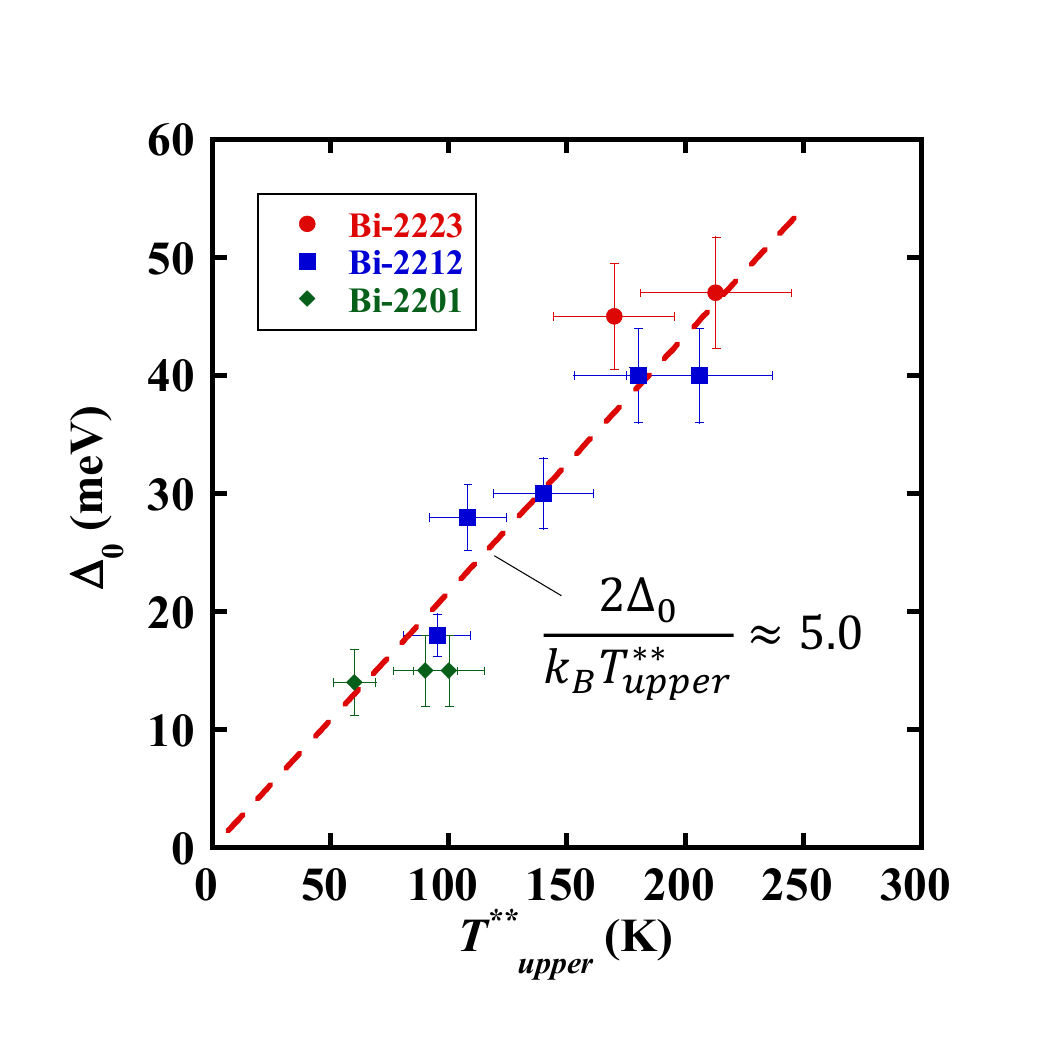}
			\caption{\label{fig4}(Color online) $\Delta_0$ versus $T^{**}_{upper}$ for Bi-2223, Bi-2212, and Bi-2201 with various hole concentrations $p$. $\Delta_0$ values are taken from Ref. \cite{Ideta25}, where $\Delta_0$ for Bi-2223 is a weighted average of the IP and OP gaps. $T^{**}_{upper}$ for Bi-2223 and Bi-2212 are obtained in this study, while those for Bi-2201 are estimated from Ref. \cite{Ando99}. A dotted line represents a linear fit to all the data points.} 
		\end{center}
	\end{figure}

To examine whether the pseudogap state is indeed independent of SCF, we perform magnetoresistance $MR$ measurements.  Figures 3(a) and 3(b) show $MR$ for optimally doped Bi‑2212 and Bi‑2223, respectively, plotted against $\tan^2 \theta_H$. In Bi‑2212, the data collapse onto a single curve above 125 K, and in Bi‑2223 above 145 K, demonstrating that the modified Kohler’s rule [Eq.(4)] holds across $T^{**}_{upper}$. Below these temperatures, an additional contribution to $MR$ emerges and grows upon cooling. We attribute this contribution to the suppression of SCF—specifically, the Aslamazov–Larkin term—by the applied magnetic field. Thus, the temperature at which $MR$ deviates from the single curve in the modified Kohler plot is identified as the onset temperature of SCF, $T_{scf}$  (for a quantitative fitting procedure for the determination of $T_{scf}$, together with complementary measurements that support our definition, see Supplemental Material \cite{Supplemental}). 

The validity of the modified Kohler's rule above $T_{scf}$ is confirmed for doping levels beyond optimal doping as well. As an example, results for underdoped Bi-2223 are shown in Fig.3(c), with additional data provided in the Supplemental Material \cite{Supplemental}. Based on these observations, the phase diagrams of Bi-2212 and Bi-2223 are constructed in Figs.3(e) and 3(f). These diagrams show that the normal state of both compounds is a strange metal in which Eqs.(1)-(4) hold, and that the SCF region is clearly distinct from the pseudogap region \cite{Grbic11}.

To clarify whether the conventional Fermi‑liquid behavior reported in Hg‑1201 \cite{Barisic13,Barisic14} is intrinsic to cuprates, we examine the same Bi‑2223 data in a Kohler plot [Fig.3(d)]. The data at 130 K and 140 K collapse onto a single curve, indicating that Kohler’s rule is satisfied within this narrow temperature window. In the same range, we observe $\rho_{ab} \propto T^2$ [Fig.1(c)] and an approximately temperature‑independent Hall coefficient [Fig.2(b)]. These behaviors can be understood by assuming that the antiferromagnetic correlation length $\xi_{AF}$ becomes nearly temperature-independent \cite{wata4}. The observed Fermi-liquid-like behavior is therefore encompassed within the broader strange‑metal phenomenology. Because Bi‑2223 contains a structurally clean inner CuO$_2$ plane, this intrinsic behavior becomes apparent. 

Here we discuss whether the pseudogap could originate from competing orders such as charge-density-wave (CDW) order. The correlation between CDW order and the pseudogap has been investigated in several Bi-based cuprates. In Bi-2201, resonant x-ray scattering (REXS) measurements show that the CDW onset temperature $T_{CDW}$ coincides with the pseudogap temperature $T^*$ obtained from Knight shift measurements \cite{Comin14}. In Bi-2212, however, Raman scattering reports that $T_{CDW}$ is slightly lower than $T^*$ \cite{Loret20}. Furthermore, in both Bi-2201 and Bi-2212, CDW order disappears in the overdoped regime \cite{Peng16,Loret20}, whereas the pseudogap persists well into overdoping \cite{Zheng05,Vishik18}. These observations indicate that CDW order is not the primary origin of the pseudogap. Instead, CDW appears as a competing order that can coexist with the pseudogap in certain doping ranges but does not track the pseudogap across the entire phase diagram.

A natural alternative origin of the pseudogap, beyond CDW order, is precursor superconductivity. Recent ARPES measurements have shown that a $d$-wave-like pairing gap persists up to $\approx 2T_c$ in the inner CuO$_2$ plane of underdoped Bi-2223 \cite{Ideta25}. The pairing gap magnitude $\Delta_0$, extrapolated from the node to the antinode, scales with $T^{**}_{upper}$, yielding $2\Delta_0/k_BT^{**}_{upper} \approx 5.0$ (Fig.4). This value is comparable to the mean‑field expectation $2\Delta_0/k_BT_c \approx 4.3$ for a $d$-wave superconductor \cite{Maki94}. These observations strongly suggest that $T^{**}_{upper}$ represents a characteristic pairing temperature \cite{Oda97,Kugler01}. Nevertheless, $T^{**}_{upper}$ is clearly distinct from $T_{scf}$ [Figs.3(e),3(f)]. 

Therefore, the pseudogap cannot be attributed to simple SCF. A more plausible interpretation is that it originates from preformed Cooper pairs in the Bardeen–Cooper–Schrieffer (BCS)–Bose–Einstein condensation (BEC) crossover regime  \cite{Nozieres85,Randeria95,Chen05,Chen24,Chen241,Shibauch20}. In this regime, the pairing interaction is sufficiently strong that the in-plane coherence length $\xi_{ab}$ becomes comparable to the average inter‑pair spacing. As a result, Cooper pairs form well above $T_c$, while global superconductivity does not emerge until these pairs undergo Bose condensation. The energy scales for pair formation and phase coherence are therefore distinct. Scanning tunneling spectroscopy supports this picture: the effective superconducting gap $\Delta_{SC}$ is reduced relative to $\Delta_0$ because the $d$-wave gap opens primarily on the Fermi arc \cite{Oda10}. 

On the other hand, theories of the crossover regime predict that the chemical potential $\mu$ should approach or fall below the band bottom \cite{Chen05}, whereas current ARPES measurements do not show such behavior \cite{Kivelson23}. We note that strong correlation effects in cuprates may pin the chemical potential even when the system is in a regime consistent with preformed pairing. Such pinning could mask the expected chemical-potential shift in ARPES. However, resolving this issue requires further experimental and theoretical work. 

Our magnetotransport measurements on Bi$_2$Sr$_2$Ca$_{n-1}$Cu$_n$O$_{2n+4+\delta} (n = 2,3)$ unambiguously demonstrate that the strong pseudogap is not a manifestation of superconducting fluctuations (SCF). Given that the strong pseudogap represents a characteristic form of pairing, this finding implies that the pseudogap in high‑$T_c$ cuprates may originate from preformed Cooper pairing in the BCS–BEC crossover regime. A broader investigation including LSCO and other cuprate families will be necessary to determine the extent to which the present conclusions apply across the entire cuprate phase diagram.

The authors acknowledge the useful discussions with H. Kontani, R. Ikeda, T. Tohyama, and A. Fujimori. This work was supported by JSPS KAKENHI Grant Numbers 25400349, 20K03849, and 23K03317. 

\section{DATA AVAILABILITY}

The data that support the findings of this article are not publicly available. The data are available from the authors upon reasonable request.


\begin{thebibliography}{100}
\bibitem{Keimer15} B. Keimer, S. A. Kivelson, M. R. Norman, S. Uchida, and J. Zaanen, From quantum matter to high-temperature
superconductivity in copper oxides, Nature \textbf{518}, 179 (2015).
\bibitem{Li10} L. Li, Y. Wang, S. Komiya, S. Ono, Y. Ando, G. D. Gu, and N. P. Ong, Diamagnetism and Cooper pairing above $T_c$ in cuprates, Phys. Rev. B \textbf{81}, 054510 (2010).
\bibitem{Wang06} Y. Wang, L. Li, and N. P. Ong, Nernst effect in high-$T_c$ superconductors, Phys. Rev. B \textbf{73}, 024510 (2006).
\bibitem{Kaiser14} S. Kaiser, C. R. Hunt, D. Nicoletti, W. Hu, I. Gierz, H. Y. Liu, M. Le Tacon, T. Loew,
D. Haug, B. Keimer, and A. Cavalleri, Optically induced coherent transport far above $T_c$ in underdoped YBa$_2$Cu$_3$O$_{6+\delta}$, Phys. Rev. B \textbf{89}, 184516 (2014).
\bibitem{Tranquada95} J. M. Tranquada, B. J. Sternlieb, J. D. Axe, Y. Nakamuya, and S. Uchida, Evidence for stripe correlations of spins and holes in copper oxide superconductors, Nature \textbf{375}, 561 (1995).
\bibitem{Ghiringhelli12} 
G. Ghiringhelli, M. L. Tacon, M. Minola, S. Blanco-Canosa, C. Mazzoli, N. B. Brookes, G. M. D. Luca, A. Frano, D. G. Hawthorn, F. He, T. Loew, M. M. Sala, D. C. Peets, M. Salluzzo, E. Schierle, R. Sutarto, G. A. Sawatzky, E. Weschke, B. Keimer, and L. Braicovich, Long-range incommensurate charge fluctuations in (Y,Nd)Ba$_2$Cu$_3$O$_{6+x}$, Science \textbf{337}, 821 (2012).
\bibitem{Daou10} R. Daou, J. Chang, D. LeBoeuf, O. Cyr-Choini\`{e}re, F. Laliber\'{e}, N. Doiron-Leyraud, B. J. Ramshaw, R. Liang, D. A. Bonn, W. N. Hardy, and L. Taillefer, Broken rotational symmetry in the pseudogap phase of a high-$T_c$ superconductor, Nature \textbf{463}, 519 (2010).
\bibitem{Timusk99} T. Timusk and B. W. Statt, The pseudogap in high temperature superconductors: an experimental survey, Rep. Prog. Phys. \textbf{62}, 61 (1999).
\bibitem{Hufner08} S. H\"{u}fner, M. A. Hossain, A. Damascelli, and G. A. Sawatzky, Two gaps make a high-temperature superconductor?, Rep. Prog. Phys. \textbf{71}, 715 (2008).
\bibitem{Kordyuk15} A. A. Kordyuk, Pseudogap from ARPES experiment: three gaps in cuprates and topological superconductivity, Low Temp. Phys. \textbf{41}, 319 (2015).
\bibitem{Vishik18} I. M. Vishik, Photoemission perspective on pseudogap, superconducting fluctuations, and chargeorder: a review of recent progress, Rep. Prog. Phys. \textbf{81}, 062501 (2018).

\bibitem{Ong91} T. R. Chien, Z. Z. Wang, and N. P. Ong, Effect of Zn impurities on the normal-state Hall angle in single-crystal YBa$_2$Cu$_{3-x}$ Zn$_x$0$_{7-\delta}$, Phys. Rev. Lett. \textbf{67}, 2088 (1991).
\bibitem{Anderson91} P. W. Anderson, Hall effect in the two-dimensional Luttinger liquid, Phys. Rev. Lett. \textbf{67}, 2092 (1991).
\bibitem{Kontani08} H. Kontani, Anomalous transport phenomena in Fermi liquids with strong magnetic fluctuations, Rep. Prog. Phys. \textbf{71}, 026501 (2008).
\bibitem{Moriya00} T. Moriya and K. Ueda, Spin fluctuations and high temperature superconductivity, Adv. Phys. \textbf{49}, 555 (2000).
%\bibitem{wata} $\rho_{ab} \propto \xi_{AF}^2 T^2$ is a result of the self‑consistent renormalization theory \cite{Moriya00}; however, since the CVC theory does not lead to qualitative changes \cite{Kontani08}, hereafter the formula will be treated as a result of the CVC theory.
\bibitem{Moriya000} T. Moriya, Y. Takahashi, and K. Ueda, Antiferromagnetic spin fluctuations and superconductivity in two-dimensional metals-A possible model for high-$T_c$ oxides, J. Phys. Soc. Jpn. \textbf{59}, 2905 (1990).
\bibitem{wata3} There has been considerable debate regarding the $T$-linear behavior of $\rho_{ab}$. See, for example, J. Zaanen, Why the temperature is high, Nature \textbf{430}, 512 (2004).

\bibitem{Ong95} J. M. Harris, Y. F. Yan, P. Matl, N. P. Ong, P. W. Anderson, T. Kimura, and K. Kitazawa, Violation of Kohler's rule in the normal-state magnetoresistanee of YBa$_2$Cu$_{3}$O$_{7-\delta}$ and La$_{2-x}$Sr$_x$CuO$_4$, Phys. Rev. Lett. \textbf{75}, 1391 (1995).
\bibitem{Kimura96} T. Kimura, S. Miyasaka, H. Takagi, K. Tamasaku, H. Eisaki, S. Uchida, K. Kitazawa, M. Hiroi, M. Sera, and N. Kobayashi, In-plane and out-of-plane magnetoresistance in La$_{2-x}$Sr$_x$CuO$_4$ single crystals, Phys. Rev. B \textbf{53}, 8733 (1996).
\bibitem{Malinowski02} A. Malinowski, Marta Z. Cieplak, S. Guha, Q. Wu, B. Kim, A. Krickser, A. Perali, K. Karpi\'nska, M. Berkowski, C. H. Shang, and P. Lindenfeld, Magnetotransport in the normal state of La$_{1.85}$Sr$_{0.15}$Cu$_{1-y}$Zn$_y$O$_4$ films, Phys. Rev. B \textbf{66}, 104512 (2002).
%\bibitem{Mackenzie98} A. W. Tyler, Y. Ando, F. F. Balakirev, A. Passner, G. S. Boebinger, A. J. Schofield, A. P. Mackenzie, and O. Laborde, High-field study of normal-state magnetotransport in Tl$_{2}$Ba$_{2}$CuO$_{6+\delta}$, Phys. Rev. B \textbf{57}, R728 (1998).
\bibitem{Hussey25} C. M. Duffy, S. J. Tu, Q. H. Chen, J. S. Zhang, A. Cuoghi, R. D. H. Hinlopen, T. Sarkar, R. L. Greene, K. Jin, and N. E. Hussey, Evidence for spin-fluctuation-mediated superconductivity in electron-doped cuprates, arXiv: 2502.13612.
\bibitem{Hussey21} J. Ayres, M. Berben, M. Čulo, Y.-T. Hsu, E. van Heumen, Y. Huang, J. Zaanen, T. Kondo, T. Takeuchi, J. R. Cooper, C. Putzke, S. Friedemann, A. Carrington, and N. E. Hussey, Incoherent transport across the strange-metal regime of overdoped cuprates, Nature \textbf{595}, 661 (2021).
\bibitem{Hussey24} J. Ayres, M. Berben, C. Duffy, R. D. H. Hinlopen, Y.-T. Hsu, A. Cuoghi, M. Leroux, I. Gilmutdinov, M. Massoudzadegan,
D. Vignolles, Y. Huang, T. Kondo, T. Takeuchi, S. Friedemann, A. Carrington, C. Proust, and N. E. Hussey, Universal correlation between $H$-linear magnetoresistance and $T$-linear resistivity in high-temperature superconductors, Nat. Commun. \textbf{15}, 8406 (2024). 

\bibitem{Kontani02} H. Kontani, Nernst coefficient and magnetoresistance in high-$T_c$ superconductors: The role of superconducting fluctuations, Phys. Rev. Lett. \textbf{89}, 237003 (2002).
\bibitem{Barisic13} N. Barišić, M. K. Chan, Y. Li, G. Yu, X. Zhao, M. Dressel, A. Smontara, and M. Greven, Universal sheet resistance and revised phase diagram of the cuprate high-temperature superconductors, Proc Natl Acad Sci USA \textbf{110}, 12235 (2013).
\bibitem{Barisic14} M. K. Chan, M. J. Veit, C. J. Dorow, Y. Ge, Y. Li, W. Tabis, Y. Tang, X. Zhao, N. Barišić, and M. Greven, In-plane magnetoresistance obeys Kohler’s rule in the pseudogap phase of cuprate superconductors, Phys. Rev. Lett. \textbf{113}, 177005 (2014).
\bibitem{Fujii02} T. Fujii, I. Terasaki, T. Watanabe, and A. Matsuda, Doping dependence of anisotropic resistivities in the trilayered superconductor Bi$_2$Sr$_2$Ca$_2$Cu$_3$O$_{10+\delta}$, Phys. Rev. B \textbf{66}, 024507 (2002).
\bibitem{Watanabe22} K. Harada, Y. Teramoto, T. Usui, K. Itaka , T. Fujii, T. Noji, H. Taniguchi,
M. Matsukawa, H. Ishikawa, K. Kindo, D. S. Dessau, and T. Watanabe, Revised phase diagram of the high-$T_c$ cuprate superconductor Pb-doped Bi$_2$Sr$_2$CaCu$_2$O$_{8+\delta}$ revealed by anisotropic transport measurements, Phys. Rev. B \textbf{105}, 085131 (2022).
%\bibitem{Watanabe24} T. Watanabe , K. Kosugi, N. Sasaki, S. Yamaguchi, T. Fujii, K. Hayama, I. Kakeya , and T. Ito, Effects of vortex and antivortex excitations in underdoped Bi$_2$Sr$_2$Ca$_2$Cu$_3$O$_{10+\delta}$ bulk single crystals, Phys. Rev. B \textbf{110}, 134509 (2024).

%\bibitem{Fujii01} T. Fujii, T. Watanabe, and A. Matsuda, Single-crystal growth of Bi$_2$Sr$ _2$Ca$_2$Cu$_3$O$_{10+\delta}$ (Bi-2223) by TSFZ method, J. Cryst. Growth \textbf{223}, 175 (2001).
%\bibitem{Adachi15} S. Adachi, T. Usui, K. Takahashi, K. Kosugi, T. Watanabe, T. Nishizaki, T. Adachi, S. Kimura, K. Sato, K. M. Suzuki, M. Fujita, K. Yamada, and T. Fujii, Single-crystal growth of underdoped Bi-2223, Physics Procedia \textbf{65}, 53 (2015).
\bibitem{Supplemental} See Supplemental Material at [URL] for details on sample preparation and the determination of the doping level, the procedures used to identify the pseudogap temperatures, 
the temperature dependence of the in-plane resistivity for underdoped Bi-2223, 
additional magnetoresistance data not shown in the main text,
and the procedures used to identify the onset temperature of superconducting fluctuations.
% estimates of the Cooper pair overlapscaling between the pseudogap values and their onset temperatures,\bibitem{Obertelli92} S. D. Obertelli, J. R. Cooper, and J. L. Tallon, Systematics in the thermoelectric power of high-$T_c$ oxides, Phys. Rev. B \textbf{46}, 14928 (1992).
%

\bibitem{Heine99} G. Heine, W. Lang, X. L. Wang, and S. X. Dou, Positive in-plane and negative out-of-plane magnetoresistance in the overdoped high-temperature superconductor Bi$_2$Sr$_2$CaCu$_2$O$_{8+x}$, Phys. Rev. B \textbf{59}, 11179 (1999).
\bibitem{Watanabe96} T. Watanabe and A. Matsuda, Magnetoresistance and high-temperature resistivity of Bi$_{2.1}$Sr$_{1.9}$Ca$_{1.0}$Cu$_2$O$_{8+\delta}$ single crystals, Physica C \textbf{263}, 313 (1996).
\bibitem{Ito93} T. Ito, K. Takenaka, and S. Uchida, Systematic deviation from $T$-linear behavior in the in-plane resistivity of YBa$_2$Cu$_3$O$_{7-y}$: Evidence for dominant spin scattering, Phys. Rev. Lett. \textbf{70}, 3995 (1993).
\bibitem{Watanabe97} T. Watanabe, T. Fujii, and A. Matsuda, Anisotropic resistivities of precisely oxygen controlled single-crystal Bi$_2$Sr$_2$CaCu$_2$O$_{8+\delta}$: Systematic study on ‘‘spin gap’’ effect, Phys. Rev. Lett. \textbf{79}, 2113 (1997).
\bibitem{Usui14} T. Usui, D. Fujiwara, S. Adachi, H. Kudo, K. Murata, H. Kushibiki, T. Watanabe, K. Kudo, T. Nishizaki, N. Kobayashi, S. Kimura, K. Yamada, T. Naito, T. Noji, and Y. Koike, Doping dependencies of onset temperatures for the pseudogap and superconductive fluctuation in Bi$_2$Sr$_2$CaCu$_2$O$_{8+\delta}$, studied from both in-plane and out-of-plane magnetoresistance measurements, J. Phys. Soc. Jpn. \textbf{83}, 064713 (2014).
\bibitem{Ando041} Y. Ando, S. Komiya, K. Segawa, S. Ono, and Y. Kurita, Electronic phase diagram of high-$T_c$ cuprate superconductors
from a mapping of the in-plane resistivity curvature, Phys. Rev. Lett. \textbf{93}, 267001 (2004).
\bibitem{Sato02} T. Sato, H. Matsui, S. Nishina, T. Takahashi, T. Fujii, T. Watanabe, and A. Matsuda, Low energy excitation and scaling in Bi$_2$Sr$_2$Ca$_{n - 1}$Cu$_n$O$_{2n+4}$ ($n$ = 1–3): Angle-resolved photoemission spectroscopy, Phys. Rev. Lett. \textbf{89}, 067005 (2002).
\bibitem{Suzuki03} Y. Yamada, K. Anagawa, T. Shibauchi, T. Fujii, T. Watanabe, A. Matsuda, and M. Suzuki, Interlayer tunneling spectroscopy and doping-dependent energy-gap structure of the trilayer superconductor Bi$_2$Sr$_2$Ca$_2$Cu$_3$O$_{10+\delta}$, Phys. Rev. B \textbf{68}, 054533 (2003).
\bibitem{Suzuki12} M. Suzuki, T. Hamatani, K. Anagawa, and T. Watanabe, Evolution of interlayer tunneling spectra and superfluid density with doping in Bi$_2$Sr$_2$CaCu$_2$O$_{8+\delta}$, Phys. Rev. B \textbf{85}, 214529 (2012).
\bibitem{Tajima24} S. Tajima, Y. Itoh, K. Mizutamari, S. Miyasaka, M. Nakajima, N. Sasaki, S. Yamaguchi, K. Harada, and T. Watanabe, Correlation between $T_c$ and the pseudogap observed in the optical spectra of high $T_c$ superconducting cuprates, J. Phys. Soc. Jpn. \textbf{93}, 103701 (2024).
\bibitem{Ando04}  Y. Ando, Y. Kurita, S. Komiya, S. Ono, and K. Segawa, Evolution of the Hall Coefficient and the Peculiar Electronic Structure
of the Cuprate Superconductors, Phys. Rev. Lett. \textbf{92}, 197001 (2004).
\bibitem{Ando00} Y. Ando, Y. Hanaki, S. Ono, T. Murayama, K. Segawa, N. Miyamoto, and S. Komiya, Carrier concentrations in Bi$_2$Sr$_{2-z}$La$_z$CuO$_{6+\delta}$ single crystals and their relation to the Hall coefficient and thermopower, Phys. Rev. B \textbf{61}, R14956 (2000).
\bibitem{Barisic19} The $T^2$ Hall-angle behavior is consistent with the findings of N. Barišić \textit{et al}., Evidence for a universal Fermi-liquid scattering rate throughout the phase diagram of the copper-oxide superconductors, New J. Phys. \textbf{21}, 113007 (2019).
%``weak''N. Barišić, M. K. Chan, M. J. Veit, C. J. Dorow, Y. Ge, Y. Li, W. Tabis, Y. Tang, G. Yu, X. Zhao, and M. Greven, Evidence for a universal Fermi-liquid scattering rate throughout the phasediagram of the copper-oxide superconductors, New J. Phys. \textbf{21}, 113007 (2019).
%\bibitem{wata1} Here, we describe the general characteristics of the temperature dependence of the Hall angle. On closer inspection, the coefficient of the $T^2$ term shows a slight change between temperatures above and below $\approx$ 200 K \cite{Watanabe22}.
\bibitem{Grbic11} The discrepancy between the region of superconducting fluctuations and the pseudogap region has also been reported in microwave absorption measurements; see M. S. Grbić \textit{et al}., Temperature range of superconducting fluctuations above $T_c$ in YBa$_2$Cu$_3$O$_{7-\delta}$ single crystals, Phys. Rev. B \textbf{83}, 144508 (2011).
%M. S. Grbić, M. Po\v{z}ek, D. Paar, V. Hinkov, M. Raichle, D. Haug, B. Keimer, N. Barišić, and A. Dul\v{c}i\'{c}, Temperature range of superconducting fluctuations above $T_c$ in YBa$_2$Cu$_3$O$_{7-\delta}$ single crystals, Phys. Rev. B \textbf{83}, 144508 (2011).
\bibitem{wata4} The reason why the antiferromagnetic correlation length $\xi_{AF}$ becomes temperature-independent is currently unknown; however, the nodal-metal state in the inner CuO$_2$ plane of the underdoped sample \cite{Ideta25} or the presence of charge-density-wave (CDW) order (or CDW fluctuations) may be responsible for this behavior.
\bibitem{Comin14} R. Comin, A. Frano, M. M. Yee, Y. Yoshida, H. Eisaki, E. Schierle, E. Weschke, R. Sutarto, F. He, A. Soumyanarayanan, Yang He, M. Le Tacon, I. S. Elfimov, Jennifer E. Hoffman, G. A. Sawatzky, B. Keimer, and A. Damascelli, Charge Order Driven by Fermi-Arc
Instability in Bi$_2$Sr$_{2-x}$La$_x$CuO$_{6+\delta}$, Science \textbf{343}, 390 (2014).
\bibitem{Loret20} B. Loret, N. Auvray, G. D. Gu, A. Forget, D. Colson, M. Cazayous, Y. Gallais, I. Paul, M. Civelli, and A. Sacuto, Universal relationship between the energy scales of the pseudogap phase, the superconducting state, and the charge-density-wave order in copper oxide superconductors, Phys. Rev. B \textbf{101}, 214520 (2020).
\bibitem{Peng16} Y. Y. Peng, M. Salluzzo, X. Sun, A. Ponti, D. Betto, A. M. Ferretti, F. Fumagalli, K. Kummer, M. Le Tacon, X. J. Zhou, N. B. Brookes, L. Braicovich, and G. Ghiringhelli, Direct observation of charge order in underdoped and optimally doped Bi$_2$(Sr,La)$_2$La$_x$CuO$_{6+\delta}$ by resonant inelastic x-ray scattering, Phys. Rev. B \textbf{94}, 184511 (2016).
\bibitem{Zheng05} Guo-qing Zheng, P. L. Kuhns, A. P. Reyes, B. Liang and C. T. Lin, Critical point and the nature of the pseudogap of single-layered copper oxide Bi$_2$Sr$_{2-x}$La$_x$CuO$_{6+\delta}$ superconductors, Phys. Rev. Lett. \textbf{94}, 047006 (2005).
%
%\bibitem{wata2} The pseudogap state in which a portion of the Fermi surface is depleted may not be a true Fermi liquid, however, we use this terminology in that the system obeys the Fermi-liquid-like behavior. 
\bibitem{Ideta25} S, Ideta, S. Adachi, T. Noji, S. Yamaguchi, N. Sasaki, S. Ishida, S. Uchida, T. Fujii, T. Watanabe, W. O. Wang, B. Moritz, T. P. Devereaux, M. Arita, C.-Y. Mou, T. Yoshida, K. Tanaka, T.-K.  Lee, and A. Fujimori, Proximity-induced nodal metal in an extremely underdoped CuO$_2$ plane in triple-layer cuprates, Nat. Commun. \textbf{16}, 9470 (2025). 

\bibitem{Ando99} Y. Ando and T. Murayama, Nonuniversal power law of the Hall scattering rate in a single-layer cuprate Bi$_2$Sr$_{2-x}$La$_x$CuO$_{6}$, Phys. Rev. B \textbf{60}, R6991 (1999).
\bibitem{Maki94} H. Won and K. Maki, d-wave superconductor as a model of high-$T_c$ superconductors, Phys. Rev. B \textbf{49}, 1397 (1994).
\bibitem{Oda97} M. Oda, K. Hoya, R. Kubota, C. Manabe, N. Momono, T. Nakano, and M. Ido, Strong pairing interactions in the underdoped region of Bi$_2$Sr$_2$CaCu$_2$O$_{8+\delta}$, Physica C \textbf{281}, 135 (1997).
\bibitem{Kugler01} M. Kugler, Ø. Fischer, Ch. Renner, S. Ono and Y. Ando, Scanning Tunneling Spectroscopy of Bi$_2$Sr$_2$CuO$_{6+\delta}$: New Evidence for the Common Origin of the Pseudogap and Superconductivity, Phys. Rev. Lett. \textbf{86}, 4911 (2001).

%\bibitem{Reber12} T. J. Reber, N. C. Plumb, Z. Sun, Y. Cao, Q. Wang, K. McElroy, H. Iwasawa, M. Arita, J. S. Wen, Z. J. Xu, G. Gu, Y. Yoshida, H. Eisaki, Y. Aiura, and D. S. Dessau, Prepairing and the ``filling'' gap in the cuprates from the tomographic density of states, Nat. Phys. \textbf{8}, 606 (2012).
%\bibitem{Kondo13} T. Kondo, A. D. Palczewski, Y. Hamaya, T. Takeuchi, J. S. Wen, Z. J. Xu, G. Gu, and A. Kaminski, Formation of gapless Fermi arcs and fingerprints of order in the pseudogap state of cuprate superconductors, Phys. Rev. Lett. \textbf{111}, 157003 (2013).
\bibitem{Nozieres85} P. Nozières and S. Schmitt-Rink, Bose condensation in an attractive fermion gas: From weak to strong coupling superconductivity, J. Low Temp. Phys. \textbf{59}, 195 (1985).
\bibitem{Randeria95} N. Trivedi and M. Randeria, Deviations from Fermi-liquid behavior above $T_c$ in 2D short coherence length superconductors, Phys. Rev. Lett. \textbf{75}, 312 (1995).
\bibitem{Chen05} Q. Chen, J. Stajic, S. Tan, and K. Levin, BCS–BEC crossover: From high temperature superconductors
to ultracold superfluids, Physics Reports \textbf{412}, 1 (2005).
\bibitem{Chen24} Q. Chen, Z. Wang, R. Boyack, S. Yang, and K. Levin, When superconductivity crosses over: from BCS to BEC, Rev. Mod. Phys. \textbf{96}, 025002 (2024).
\bibitem{Chen241} Q. Chen, Z. Wang, R. Boyack, and K. Levin,, Test for BCS-BEC crossover in the cuprate superconductors, npj quantum materials \textbf{9}, 27 (2024).
\bibitem{Shibauch20} T. Shibauchi, T. Hanaguri, and Y. Matsuda, Exotic Superconducting States in FeSe-based Materials, J. Phys. Soc. Jpn. \textbf{89}, 102002 (2020).
\bibitem{Oda10} T. Kurosawa, T. Yoneyama, Y. Takano, M. Hagiwara, R. Inoue, N. Hagiwara, K. Kurusu, K. Takeyama,
N. Momono, M. Oda, and M. Ido, Large pseudogap and nodal superconducting gap in Bi$_2$Sr$_{2-x}$La$_x$CuO$_{6+\delta}$ and Bi$_2$Sr$_2$CaCu$_2$O$_{8+\delta}$: Scanning tunneling microscopy and spectroscopy, Phys. Rev. B \textbf{81}, 094519 (2010).
%\bibitem{Suzuki00} M. Suzuki and T. Watanabe, Discriminating the superconducting gap from the pseudogap in Bi$_2$Sr$_2$CaCu$_2$O$_{8+\delta}$ by interlayer tunneling spectroscopy, Phys. Rev. Lett. \textbf{85}, 4787 (2000).
%\bibitem{Matsuda99} A. Matsuda, S. Sugita, and T. Watanabe, Temperature and doping dependence of the Bi$_{2.1}$Sr$_{1.9}$CaCu$_2$O$_{8+\delta}$ pseudogap and superconducting gap, Phys. Rev. B \textbf{60}, 1377 (1999).
%\bibitem{Ando07} S. Ono, S. Komiya, and Y. Ando, Strong charge fluctuations manifested in the high-temperature Hall coefficient of high-$T_c$ cuprates, Phys. Rev. B \textbf{75}, 024515 (2007).
%\bibitem{Norman98} M. R. Norman, H. Ding, M. Randeria, J. C. Campuzano, T. Yokoya, T. Takeuchi, T. Takahashi, T. Mochiku, K. Kadowaki, P. Guptasarma, and D. G. Hinks, Destruction of the Fermi surface in underdoped high-$T_c$ superconductors, Nature \textbf{392}, 157 (1998).
%\bibitem{Adachi151} S. Adachi, T. Usui, Y. Ito, H. Kudo, H. Kushibiki, K. Murata, T. Watanabe, K. Kudo, T. Nishizaki, N. Kobayashi, S. Kimura, M. Fujita, K. Yamada, T. Noji, Y. Koike, and T. Fujii, Unscaling superconducting parameters with $T_c$ for Bi-2212 and Bi-2223: a magnetotransport study in the superconductive fluctuation regime, J. Phys. Soc. Jpn. \textbf{84}, 024706 (2015).
%\bibitem{Ikeda91} R. Ikeda, T. Ohmi, and T. Tsuneto, Theory of broad resistive transition in high temperature superconductors under magnetic field, J. Phys. Soc. Jpn. \textbf{60}, 1051 (1991).
\bibitem{Kivelson23} J. Sous, Y. He, and S. A. Kivelson, Absence of a BCS-BEC crossover in the cuprate superconductors, npj quantum materials \textbf{8}, 25 (2023).

%\bibitem{Ando991} Y. Abe, Y. Ando, J. Takeya, H. Tanabe, T. Watauchi, I. Tanaka, and H. Kojima, Normal-state magnetotransport in La$_{1.905}$Ba$_{0.095}$CuO4 single crystals, Phys. Rev. B \textbf{59}, 14753 (1999).

%\bibitem{Barisic19} N. Barišić, M. K. Chan, M. J. Veit, C. J. Dorow, Y. Ge, Y. Li, W. Tabis, Y. Tang, G. Yu, X. Zhao, and M. Greven, Evidence for a universal Fermi-liquid scattering rate throughout the phasediagram of the copper-oxide superconductors, New J. Phys. \textbf{21}, 113007 (2019).
%\bibitem{Ogata08} M. Ogata and H. Fukuyama, The $t–J$ model for the oxide high-$T_c$ superconductors, Rep. Prog. Phys. \textbf{71}, 036501 (2008).
%\bibitem{Kivelson95} V. J. Emery and S. A. Kivelson, Importance of phase fluctuations in superconductors with small superfluid density, Nature \textbf{374}, 434 (1995).


%\bibitem{Uemura89} Y. J. Uemura, G. M. Luke, B. J. Sternlieb, J. H. Brewer, J. F.
%Carolan, W. N. Hardy, R. Kadono, J. R. Kempton, R. F. Kiefl,
%S. R. Kreitzman, P. Mulhern, T. M. Riseman, D. L. Williams,
%B. X. Yang, S. Uchida, H. Takagi, J. Gopalakrishnan, A. W.
%Sleight, M. A. Subramanian, C. L. Chien, M. Z. Cieplak,
%G. Xiao, V. Y. Lee, B. W. Statt, C. E. Stronach, W. J. Kossler,
%and X. H. Yu, Phys. Rev. Lett. \textbf{62}, 2317 (1989).
%\bibitem{Kivelson95} V. J. Emery and S. A. Kivelson, Nature \textbf{374}, 434 (1995).
%\bibitem{Franz98} M. Franz and A. J. Millis, Phys. Rev. B \textbf{58}, 14572 (1998).
%\bibitem{Ong00} Z. A. Xu, N. P. Ong, Y. Wang, T. Kakeshita, and S. Uchida,
%Nature \textbf{406}, 486 (2000).
%\bibitem{Ong05} Y. Wang, L. Li, M. J. Naughton, G. D. Gu, S. Uchida, and N. P. Ong, Phys. Rev. Lett. \textbf{95}, 247002 (2005).
%\bibitem{Kosterlitz73} J. M. Kosterlitz and D. J. Thouless, J. Phys. C \textbf{6}, 1181 (1973).
%\bibitem{Beasley79} M. R. Beasley, J. E. Mooij, and T. P. Orlando, Phys. Rev. Lett. \textbf{42}, 1165 (1979).
%\bibitem{Franz07} M. Franz, Nature Phys. \textbf{3}, 686 (2007).
%\bibitem{Matsuda92} Y. Matsuda, S. Komiyama, T. Terashima, K. Shimura, and Y. Bando, Phys. Rev. Lett. \textbf{69}, 3228 (1992).
%\bibitem{Hetel07} I. Hetel, T. R. Lemberger, and M. Randeria, Nature Phys. \textbf{3}, 700 (2007).
%\bibitem{Yu22} A. B. Yu, Z. Huang,W. Peng, H. Li, C. T. Lin, X. F. Zhang, and L. X. You, Appl. Phys. Lett. \textbf{120}, 072601 (2022).
%\bibitem{Tranquada07} Q. Li, M. H\"{u}cker, G. D. Gu, A. M. Tsvelik, and J. M. Tranquada, Phys. Rev. Lett. \textbf{99}, 067001 (2007).
%\bibitem{Kitano06} H. Kitano, T. Ohashi, A. Maeda, and I. Tsukada, Phys. Rev. B \textbf{73}, 092504 (2006).
%\bibitem{Matsuda93} Y. Matsuda, S. Komiyama, T. Onogi, T. Terashima, K. Shimura, and Y. Bando, Phys. Rev. B \textbf{48}, 10498 (1993).
%\bibitem{Hikami80} S. Hikami and T. Tsuneto, Prog. Theor. Phys. \textbf{63}, 387 (1980).
%\bibitem{Mukuda12} H. Mukuda, S. Shimizu, A. Iyo, , and Y. Kitaoka, J. Phys. Soc. Jpn. \textbf{81}, 011008 (2012).
%\bibitem{Kunisada20} S. Kunisada, S. Isono, Y. Kohama, S. Sakai, C. Bareille, S. Sakuragi, R. Noguchi, K. Kurokawa, K. Kuroda, Y. Ishida, S. Adachi, R. Sekine, T. K. Kim, C. Cacho, S. Shin, T. Tohyama, K. Tokiwa, and T. Kondo, Science \textbf{369}, 833 (2020).
%\bibitem{Iye10} T. Iye, T. Nagatochi, R. Ikeda, and A. Matsuda, J. Phys. Soc. Jpn. \textbf{79}, 114711 (2010).
%\bibitem{Nomura19} Y. Nomura, R. Okamoto, T. A. Mizuno, S. Adachi, T. Watanabe, M. Suzuki, and I. Kakeya, Phys. Rev. B \textbf{100}, 144515 (2019).
%\bibitem{Fujii01} T. Fujii, T. Watanabe, and A. Matsuda, J. Cryst. Growth \textbf{223}, 175 (2001).


%\bibitem{Aslamasov68} L. G. Aslamasov and A. I. Larkin, Phys. Lett. A \textbf{26}, 238 (1968).
%\bibitem{Minnhagen87} P. Minnhagen, Rev. Mod. Phys. \textbf{59}, 1001 (1987).
%\bibitem{Halperin79} B. I. Halperin and D. R. Nelson, J. Low Temp. Phys. \textbf{36}, 599 (1979).
%\bibitem{Blatter94} G. Blatter, M. Y. Feigel’man, Y. B. Geshkenbein, A. I. Larkin, and V. M. Vinokur, Rev. Mod. Phys. \textbf{66}, 1125 (1994).
%\bibitem{Schilling93} A. Schilling, R. Jin, J. D. Guo, and H. R. Ott, Phys. Rev. Lett. \textbf{71}, 1899 (1993).
%\bibitem{Mackenzie93} A. P. Mackenzie, S. R. Julian, G. G. Lonzarich, A. Carrington, S. D. Hughes, R. S. Liu, and D. C. Sinclair, Phys. Rev. Lett. \textbf{71}, 1238 (1993).
%\bibitem{Li07} L. Li, J. G. Checkelsky, S. Komiya, Y. Ando, and N. P. Ong, Nature Physics \textbf{3}, 311 (2007).
%\bibitem{Hsu21} Y.-T. Hsu, M. Hartstein, A. J. Davies, A. J. Hickey, M. K. Chan, J. Porras, T. Loew, S. V. Taylor, H. Liu, A. G. Eaton, M. L. Tacon, H. Zuo, J. Wang, Z. Zhu, G. G. Lonzarich, B. Keimer, N. Harrison, and S. E. Sebastian, Proc. Natl. Acad. Sci. U.S.A. \textbf{118}, 2021216118 (2021).
%\bibitem{Vinokur90} V. M. Vinokur, P. H. Kes, and A. E. Koshelev, Physica C \textbf{168}, 29 (1990).
%\bibitem{Piriou08} A. Piriou, Y. Fasano, E. Giannini, and O. Fischer, Phys. Rev. B \textbf{77}, 184508 (2008).
%\bibitem{Blatter03} G. Blatter and V. B. Geshkenbein, ”Vortex matter” in \textit{The
%Physics of Superconductors, vol. I}, K. H. Bennemann, J. B. Ketterson,
%Eds. (Springer, Berlin, Germany, 2003) , pp. 725.
%\bibitem{Cohen97} L. F. Cohen and H. J. Jensen, Rep. Prog. Phys. \textbf{60}, 1581 (1997).
%\bibitem{Korshunov90} S. E. Korshunov, Europhys. Lett. \textbf{11(8)}, 757 (1990).
%\bibitem{Geshkenbein98} V. B. Geshkenbein, L. B. Ioffe, and A. J. Millis, Phys. Rev. Lett. \textbf{80}, 5778 (1998).
%\bibitem{Ikeda06} R. Ikeda, Phys. Rev. B \textbf{74}, 054510 (2006).
%\bibitem{Zhou03} X. J. Zhou, T. Yoshida, A. Lanzara, P. V. Bogdanov, S. A. Kellar, K. M. Shen, W. L. Yang, F. Ronning, T. Sasagawa, T. Kakeshita, T. Noda, H. Eisaki, S. Uchida, C. T. Lin, F. Zhou, J.W. Xiong,W. X. Ti, Z. X. Zhao, A. Fujimori, Z. Hussain, and Z.-X. Shen, Nature \textbf{423}, 398 (2003).
%\bibitem{Yamada03} Y. Yamada, K. Anagawa, T. Shibauchi, T. Fujii, T. Watanabe, A. Matsuda, and M. Suzuki, Phys. Rev. B \textbf{68}, 054533 (2003).

%\bibitem{Watanabe00} T. Watanabe, T. Fujii, and A. Matsuda, Phys. Rev. Lett. \textbf{84}, 5848 (2000).
%\bibitem{Martin89} S. Martin, A. T. Fiory, R. M. Fleming, G. P. Espinosa, , and A. S. Cooper, Phys. Rev. Lett. \textbf{62}, 677 (1989).
%\bibitem{Pan01} S. H. Pan, J. P. O’Neal, R. L. Badzey, C. Chamon, H. Ding, J. R. Engelbrecht, Z. Wang, H. Eisaki, S. Uchida, A. K. Gupta, K.-W. Ng, E. W. Hudson, K. M. Lang, and J. C. Davis, Nature \textbf{413}, 282 (2001).
%\bibitem{Lang02} K. M. Lang, V. Madhavan, J. E. Hoffman, E. W. Hudson, H. Eisaki, S. Uchida, and J. C. Davis, Nature \textbf{415}, 412 (2002).
%\bibitem{Gomes07} K. K. Gomes, A. N. Pasupathy, A. Pushp, S. Ono, Y. Ando, and A. Yazdani, Nature \textbf{447}, 569 (2007).
%\bibitem{Kasai09} T. Kasai, H. Nakajima, T. Fujii, I. Terasaki, T. Watanabe, H. Shibata, and A. Matsuda, Physica C \textbf{469}, 1016 (2009).
%\bibitem{Hamidian16} M. H. Hamidian, S. D. Edkins, S. H. Joo, A. Kostin, H. Eisaki, S. Uchida, M. J. Lawler, E.-A. Kim, A. P. Mackenzie, K. Fujita, J. Lee, and J. C. Davis, Nature \textbf{532}, 343 (2016).
%\bibitem{Du20} Z. Du, H. Li, S. H. Joo, E. P. Donoway, J. Lee, J. C. Davis, G. Gu, P. D. Johnson, and K. Fujita, Nature \textbf{580}, 65 (2020).
%\bibitem{Semba00} K. Semba, M. Mukaida, and A. Matsuda, in \textit{Proceedings of the Mass and Charge Transport in Inorganic Materials: Fundamentals to Devices, Part A, Venezia, Italy, 2000}, edited by P. Vincenzini and V. Buscaglia (Techna Srl, 2000, ISBN:88- 86538-30-8) , 121 (2000).
%\bibitem{Nozieres85} P. Nozieres and S. Schmitt-Rink, Bose condensation in an attractive fermion gas: from weak to strong coupling superconductivity, J. Low Temp. Phys. \textbf{59}, 195 (1985).
\end{thebibliography}
\end{document}

%\preprint{APS/123-QED}

\title{Discriminating superconducting fluctuations from the pseudogap in Bi$_2$Sr$_2$Ca$_{n-1}$Cu$_n$O$_{2n+4+\delta} (n = 2,3)$:  A magnetotransport study} % Force line breaks with \\

\author{Shunpei Yamaguchi}
\affiliation{Graduate School of Science and Technology, Hirosaki University, Hirosaki, Aomori, 036-8561 Japan}
\author{Nae Sasaki}
\affiliation{Graduate School of Science and Technology, Hirosaki University, Hirosaki, Aomori, 036-8561 Japan}
\author{Shintaro Adachi}
\affiliation{Graduate School of Science and Technology, Hirosaki University, Hirosaki, Aomori, 036-8561 Japan}
\affiliation{Faculty of Engineering/Graduate School of Engineering, Kyoto University of Advanced Science (KUAS), 615-8577 Japan}	
\author{Keiichi Harada}
\affiliation{Graduate School of Science and Technology, Hirosaki University, Hirosaki, Aomori, 036-8561 Japan}
\author{Yuki Teramoto}
\affiliation{Graduate School of Science and Technology, Hirosaki University, Hirosaki, Aomori, 036-8561 Japan}
\author{Shintaro Matsuda}
\affiliation{Graduate School of Science and Technology, Hirosaki University, Hirosaki, Aomori, 036-8561 Japan}
\author{Tomohiro Usui}
\affiliation{Graduate School of Science and Technology, Hirosaki University, Hirosaki, Aomori, 036-8561 Japan}

	%\affiliation{Graduate School of Science and Technology, Hirosaki University, 3 Bunkyo, Hirosaki, 036-8561 Japan}
	%\author{Mihaly M. Dobroka$^1$}
	%\affiliation{Graduate School of Science and Technology, Hirosaki University, 3 Bunkyo, Hirosaki, 036-8561 Japan}
%	\author{Shintaro Adachi}
% \affiliation{Nagamori Institute of Actuators, Kyoto University of Advanced Science (KUAS), Kyoto 615-8577, Japan}
        \author{Takenori Fujii}
\affiliation{Cryogenic Research Center, University of Tokyo, Bunkyo, Tokyo 113-0032, Japan}
\author{Takashi Noji}
\affiliation{Graduate School of Engineering, Tohoku University, Sendai 980-8579, Japan}			
	\author{Itsuhiro Kakeya}
\affiliation{Department of Electronic Science and Engineering, Kyoto University, Kyoto 615-8510, Japan}
\author{Haruka Taniguchi}
\affiliation{Graduate School of Engineering, Iwate University, Morioka 020-8551, Japan}
	%\affiliation{Department of Electronic Science and Engineering, Kyoto University, Kyoto 615-8510, Japan}
	\author{Michiaki Matsukawa}
\affiliation{Graduate School of Engineering, Iwate University, Morioka 020-8551, Japan}
	%\affiliation{Institute for Solid State Physics, University of Tokyo, 5-1-5 Kashiwanoha, Kashiwa, Chiba 277-8581, Japan}
	%\author{Koichi Kindo$^3$}
	%\affiliation{Institute for Solid State Physics, University of Tokyo, 5-1-5 Kashiwanoha, Kashiwa, Chiba 277-8581, Japan }
\author{Atsushi Miyake}
\affiliation{Institute for Solid State Physics, University of Tokyo, Kashiwa, Chiba 277-8581, Japan}
	\author{Hajime Ishikawa}
\affiliation{Institute for Solid State Physics, University of Tokyo, Kashiwa, Chiba 277-8581, Japan}
        \author{Koichi Kindo}
\affiliation{Institute for Solid State Physics, University of Tokyo, Kashiwa, Chiba 277-8581, Japan}
        %\author{Toshimitsu Ito}
%\affiliation{Research Institute for Advanced Electronics and Photonics, National Institute of Advanced Industrial Science and Technology (AIST), Higashi 1-1-1, Tsukuba, Ibaraki 305-8565, Japan}
\author{Takao Watanabe}
\email{E-mail address: watanabe.takao@nihon-u.ac.jp}
%\email{Present address: Physics Department, College of Engineering, Nihon University, Fukushima 963-8642, Japan. E-mail address: 
%watanabe.takao@nihon-u.ac.jp}
\affiliation{Graduate School of Science and Technology, Hirosaki University, Hirosaki, Aomori, 036-8561 Japan}
\affiliation{Physics Department, College of Engineering, Nihon University, Fukushima 963-8642, Japan}
\affiliation{Department of Advanced Materials Science, University of Tokyo, Kashiwa, Chiba 277-8561, Japan}
%        \author{Shojiro Kimura}
% \affiliation{Institute for Materials Research, Tohoku University, 2-1-1 Katahira, Aoba-ku, Sendai, 980-8577 Japan}
%	\author{Ken Hayama}
%\affiliation{Department of Electronic Science and Engineering, Kyoto University, Kyoto 615-8510, Japan}

	%\affiliation{Institute for Solid State Physics, University of Tokyo, 5-1-5 Kashiwanoha, Kashiwa, Chiba 277-8581, Japan}
	%\author{Koichi Kindo$^3$}
	%\affiliation{Institute for Solid State Physics, University of Tokyo, 5-1-5 Kashiwanoha, Kashiwa, Chiba 277-8581, Japan }
	%\author{Hajime Ishikawa}
%\affiliation{Institute for Solid State Physics, University of Tokyo, Kashiwa, Chiba 277-8581, Japan}
        %\author{Koichi Kindo}
%\affiliation{Institute for Solid State Physics, University of Tokyo, Kashiwa, Chiba 277-8581, Japan}
       % \author{Daniel S. Dessau}
% \affiliation{Department of Physics, University of Colorado at Boulder, Boulder, CO 80309, USA}

	%\affiliation{Institute for Materials Research, Tohoku University, 2-1-1 Katahira, Aoba-ku, Sendai, 980-8577 Japan}
%	\author{Takao Watanabe}
%\email{twatana@hirosaki-u.ac.jp}
%\affiliation{Graduate School of Science and Technology, Hirosaki University, Hirosaki, Aomori, 036-8561 Japan}
	%\affiliation{Graduate School of Science and Technology, Hirosaki University, 3 Bunkyo, Hirosaki, 036-8561 Japan}
%$^\star$\thanks{email: twatana@hirosaki-u.ac.jp}

	%\affiliation{Institute for Solid State Physics, University of Tokyo, 5-1-5 Kashiwanoha, Kashiwa, Chiba 277-8581, Japan$^3$}

%\author{Aaa Bee}

%\affiliation{Hirosaki University, Japan}


\date{\today}% It is always \today, today,
\maketitle

\section*{A. details of sample preparation and determination of the doping level}

\begin{table}

   \caption{Summary of annealing conditions, $T_c$, and doping levels ($p$) for Bi-2212.}
   \label{tab1}
   \vspace*{0.5cm}

\begin{tabular}{cccc}
\hline
\hline
Annealing Conditions & $T_c$ & $p$ \\
\hline
600 $^\circ$C, 24 h in O$_2$ 2 Pa  & 72 K & 0.126 \\
600 $^\circ$C, 12 h in O$_2$ 2 Pa  & 89 K & 0.179 \\
600 $^\circ$C, 12 h in O$_2$ 100 Pa  & 83 K & 0.194 \\
600 $^\circ$C, 18 h in O$_2$ 1 $\%$  & 76 K & 0.206 \\
400 $^\circ$C, 200 h in O$_2$ 1 atm  & 64 K & 0.220 \\
\hline
\hline
\end{tabular} 
\end{table}

Single crystals of Pb-doped Bi-2212, Bi$_{1.6}$Pb$_{0.4}$Sr$_2$CaCu$_2$O$_{8+\delta}$ 
(nominal composition Bi$_{1.6}$Pb$_{0.6}$Sr$_2$CaCu$_2$O$_{8+\delta}$) were grown in air using the traveling-solvent floating-zone (TSFZ) method \cite{Watanabe22}. 
These crystals were used for optimally doped to overdoped samples. 
For the underdoped sample, single crystals of Bi$_2$Pb$_{0.4}$Sr$_{1.8}$CaCu$_2$O$_{8+\delta}$ 
(nominal composition Bi$_2$Pb$_{0.6}$Sr$_{1.8}$CaCu$_2$O$_{8+\delta}$) were also grown using the TSFZ method. 
The crystals were annealed under various oxygen partial pressures $P_{\mathrm{O2}}$ 
(2 Pa $\le P_{\mathrm{O2}} \le$ 1 atm) at 400–600$^\circ$C for sufficiently long durations to homogeneously tune the doping level. 

Table I summarizes the annealing conditions, $T_c$, and the corresponding hole concentrations $p$. 
$T_c$ was determined from the onset of zero resistivity. 
The doping level $p$ of Bi-2212 was obtained using the empirical relation \cite{Obertelli92},

\begin{equation}
T_c/T_c^{\mathrm{max}} = 1 - 82.6(p - 0.16)^2,
%T_c/T_c^{max} = 1 - 82.6(p - 0.16)^2,
\end{equation}
%with $T_c^{max}$ = 91.7 K and 76.0 K for  Bi$_{1.6}$Pb$_{0.4}$Sr$_2$CaCu$_2$O$_{8+\delta}$ and Bi$_{2}$Pb$_{0.4}$Sr$_{1.8}$CaCu$_2$O$_{8+\delta}$, respectively.
with $T_c^{\mathrm{max}} = 91.7$ K and 76.0 K for 
Bi$_{1.6}$Pb$_{0.4}$Sr$_2$CaCu$_2$O$_{8+\delta}$ and 
Bi$_2$Pb$_{0.4}$Sr$_{1.8}$CaCu$_2$O$_{8+\delta}$, respectively. Equation (1) is known to hold reliably in the doping range studied here (0.126 $\le p \le$ 0.220). Because this formula provides a well-defined and single-valued mapping between $T_c$ and $p$, the uncertainty in $p$ is smaller than the symbol size in the phase diagrams. Therefore, error bars for Bi-2212 are not visible but are implicitly included within the plotting marks. 

High-quality Bi-2223 single crystals were grown using the TSFZ method \cite{Fujii01,Adachi15}. 
The feed-rod composition was Bi-rich (Bi:Sr:Ca:Cu = 2.2–2.25 : 1.9–2.0 : 2.0 : 2.9–3.0), 
the growth atmosphere was an O$_2$–Ar mixture containing 10\% oxygen, 
and the growth rate was set to 0.05 mm/h. 
The obtained crystals were annealed under various oxygen partial pressures 
(2 Pa $\le P_{\mathrm{O2}} \le$ 400 atm) at 400–600$^\circ$C. 
Table II summarizes the annealing conditions, $T_c$, and the corresponding doping levels.

In contrast to Bi-2212, Bi 2223 does not follow the empirical relation above \cite{Fujii02}. Thus, we used doping levels determined by angle-resolved photoemission spectroscopy (ARPES) \cite{Ideta25}, where the hole concentration of the inner and outer CuO$_2$ planes ($p_{\mathrm{IP}}$ and $p_{\mathrm{OP}}$) were obtained directly from the Fermi-surface volume. The overall doping level was evaluated using the weighted average,  

\begin{equation}
p = (p_{\mathrm{IP}} + 2p_{\mathrm{OP}})/3.
\end{equation}

The samples used in our transport measurements were annealed under the same conditions as those used in the ARPES study, ensuring consistency. However, the ARPES-derived doping values include experimental uncertainties, and additional small sample-to-sample variations exist. Therefore, the error bars for Bi-2223 in our phase diagrams are set slightly larger than those reported in the ARPES study, reflecting both measurement uncertainty and sample variation.

\begin{table}

   \caption{Summary of annealing conditions, $T_c$, and doping levels ($p$) for Bi-2223.}
   \label{tab1}
   \vspace*{0.5cm}

\begin{tabular}{cccc}
\hline
\hline
Annealing Conditions & $T_c$ & $p$ \\
\hline
500 $^\circ$C, 1 h in O$_2$ 2 Pa  & 75 K & 0.09 \\
600 $^\circ$C, 5 h in O$_2$ 100 Pa  & 95 K & 0.10 \\
500 $^\circ$C, 50 h in O$_2$ 1 atm  & 104 K & 0.17 \\
400 $^\circ$C, 50 h in O$_2$ 400 atm  & 105 K & 0.20 \\
\hline
\hline
\end{tabular} 
\end{table}

\begin{figure}[t]
		\begin{center}
			
			\includegraphics[width=85mm]{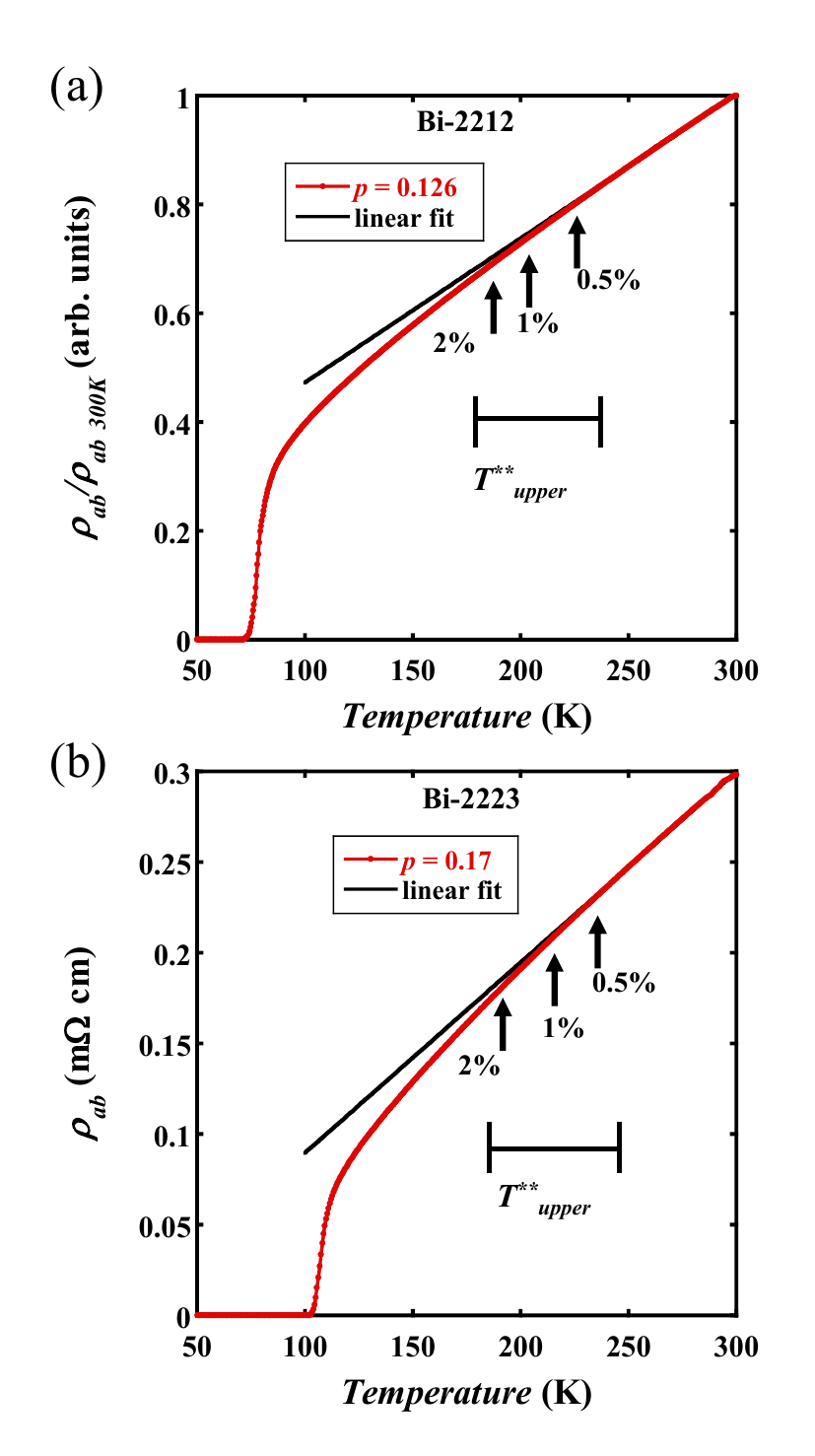}
			\caption{(Color online)
Sensitivity analysis of $T^{**}_{\mathrm{upper}}$ obtained by varying the deviation threshold (0.5$\%$, 1$\%$, and 2$\%$ for (a) Bi-2212 ($p$ = 0.126) and (b) Bi-2223 ($p$ = 0.17). The error bars indicate the uncertainty in $T^{**}_{\mathrm{upper}}$.
}
\label{figS1}

		\end{center}
	\end{figure}

\begin{figure}[t]
		\begin{center}
			
			\includegraphics[width=80mm]{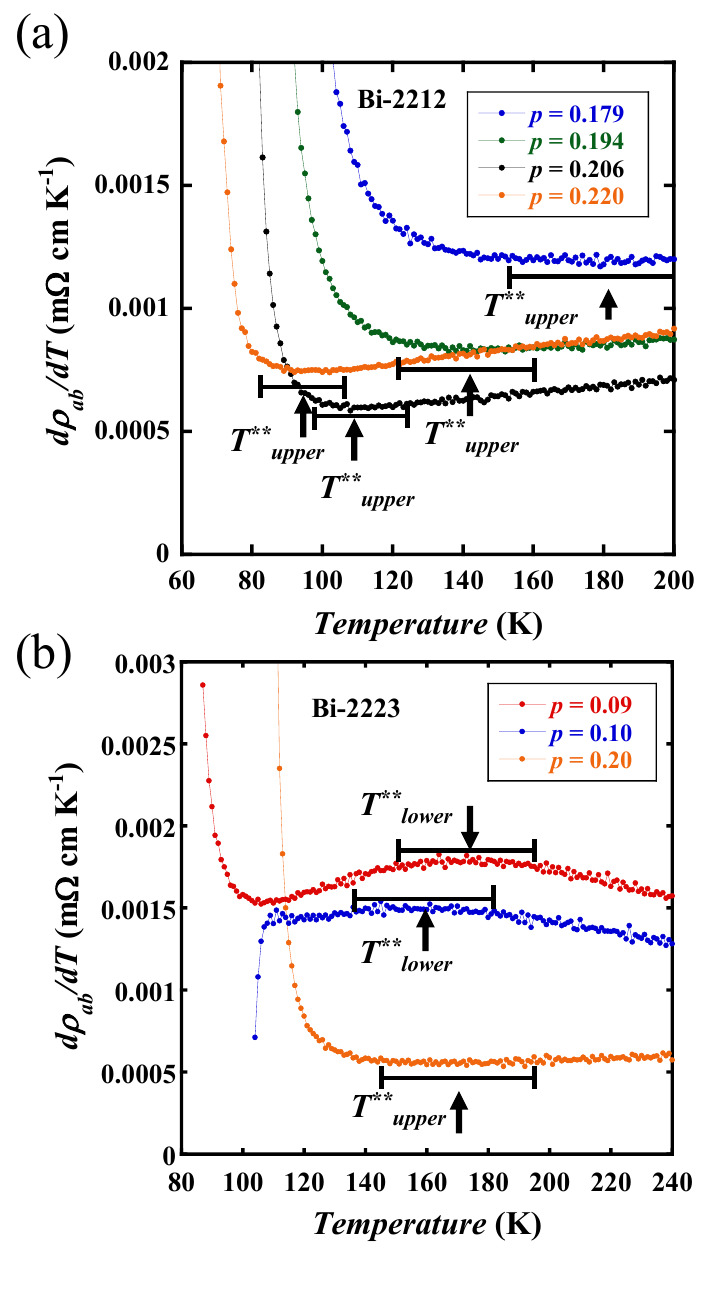}
			\caption{(Color online)
(a) $d\rho_{ab}/dT$ versus $T$ for Bi-2212 ($p$ = 0.179, 0.194, 0.206, and 0.220). 
(b) $d\rho_{ab}/dT$ versus $T$ for Bi-2223 ($p$ = 0.09, 0.10, and 0.20). 
Arrows indicate the pseudogap temperatures $T^{**}_{\mathrm{upper}}$ and $T^{**}_{\mathrm{lower}}$. The error bars indicate the uncertainty in $T^{**}_{\mathrm{upper}}$ and $T^{**}_{\mathrm{lower}}$.
}
\label{figS1}
			
		\end{center}
	\end{figure}

%An upward arrow denotes the pseudogap opening temperature $T^{**}_{upper}$, while downward arrows denote the characteristic temperature for the pseudogap, 

\begin{figure*}[t]
		\begin{center}
			
			\includegraphics[width=180mm]{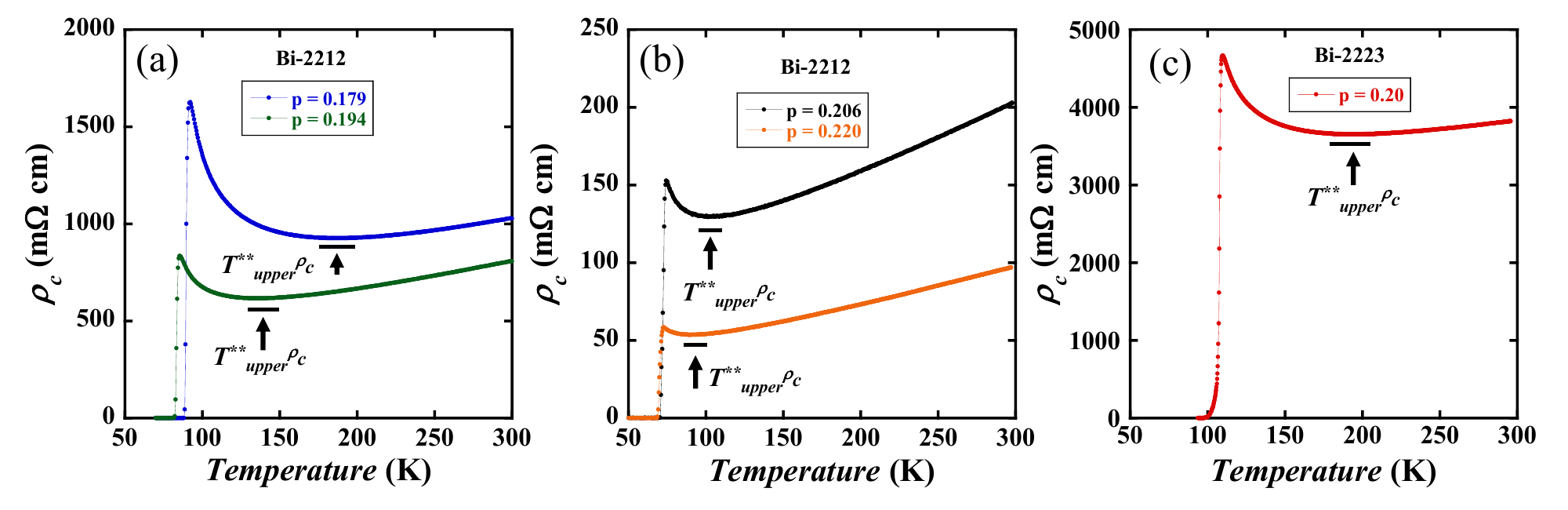}
			\caption{(Color online)
Out of plane resistivity $\rho_{c}(T)$ for (a) Bi-2212 with $p$ = 0.179 and 0.194, (b) Bi-2212 with $p$ = 0.206 and 0.220, and (c) Bi-2223 with $p$ = 0.20. Arrows mark the pseudogap onset temperatures $T_{\mathrm{upper}}^{**\quad\rho_{c}}$. Horizontal black bars indicate the temperature ranges over which $\rho_{c}$ remains effectively constant within experimental uncertainty; these ranges fall within the error bars used in the phase diagrams (see also Figs. 2 and 4).
}
\label{figS1}
			
		\end{center}
	\end{figure*}

\begin{figure*}[t]
		\begin{center}
			
			\includegraphics[width=150mm]{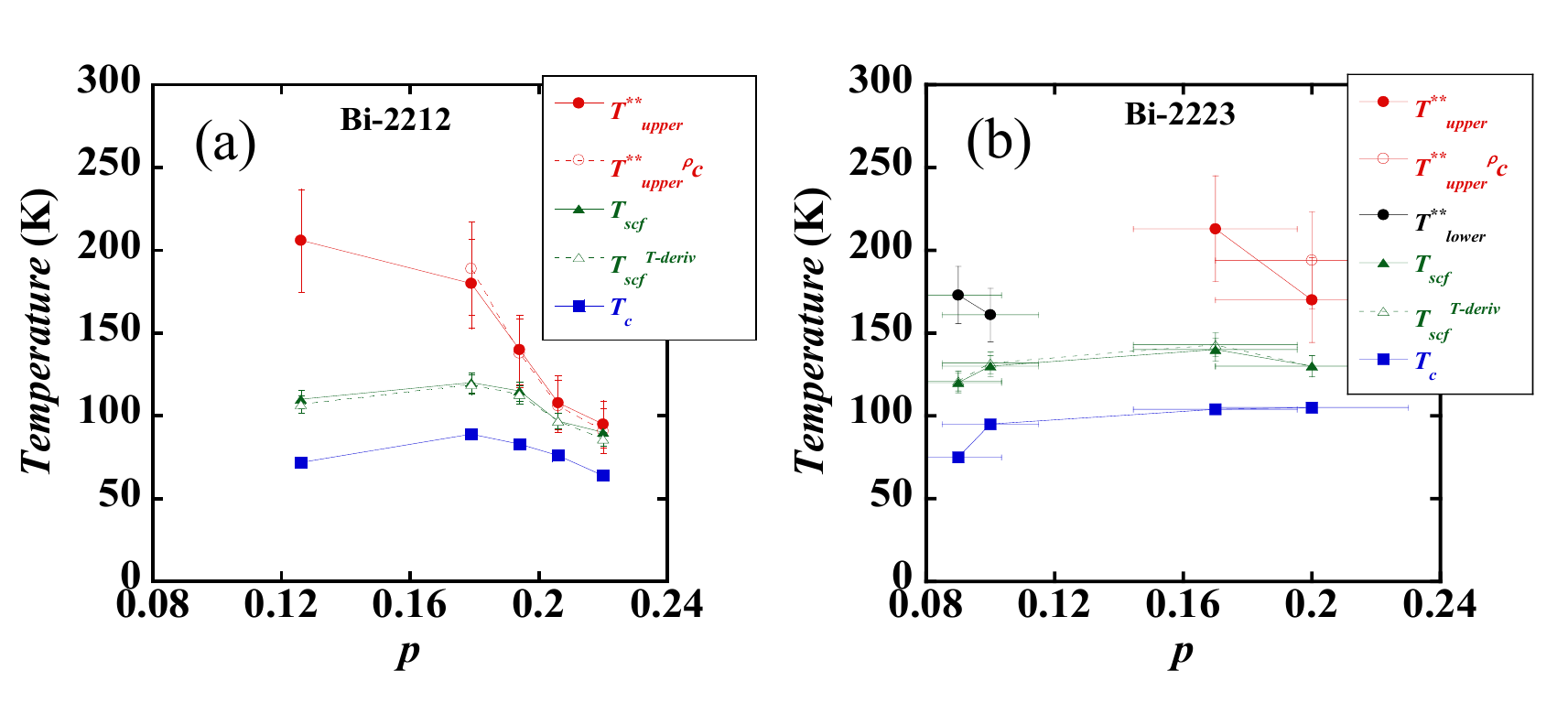}
			\caption{(Color online)
Comparison of onset temperatures for the pseudogap and superconducting fluctuations obtained from different methods for (a) Bi-2212 and (b) Bi-2223.
}
\label{figS1}

		\end{center}
	\end{figure*}

\begin{figure}[t]
		\begin{center}
			
			\includegraphics[width=85mm]{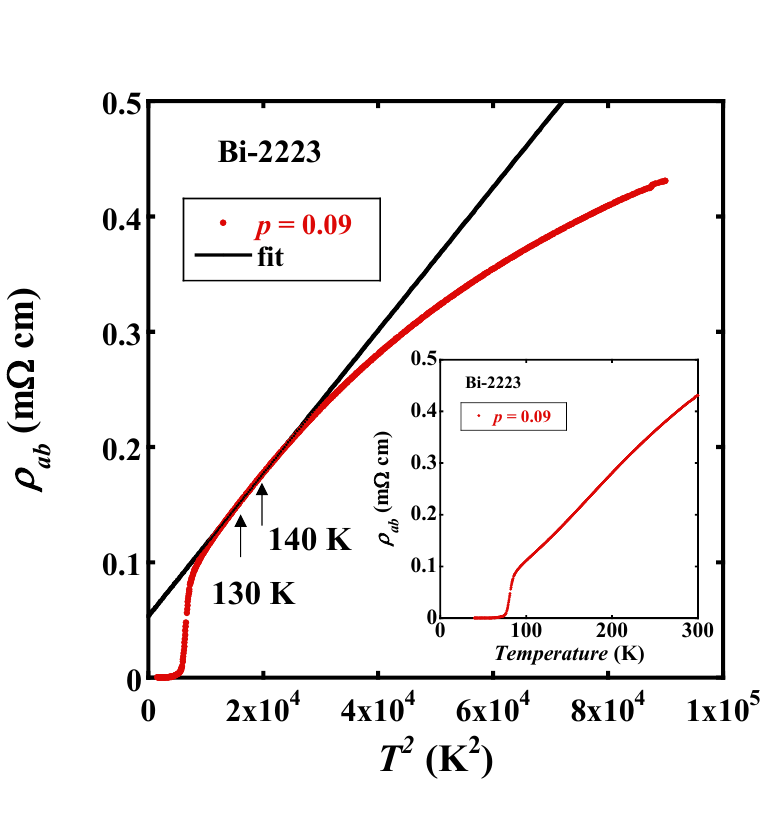}
			\caption{(Color online)
In-plane resistivity $\rho_{ab}(T)$ of Bi-2223 with $p$ = 0.09 plotted versus $T^2$. 
The solid straight line is a fit to $\rho_{ab} = AT^2 + B$, with 
$A = 6.20 \times 10^{-6}$ m$\Omega$ cm K$^{-2}$ and 
$B = 0.053$ m$\Omega$ cm. 
Inset: $\rho_{ab}(T)$ plotted versus $T$ for the same data.
}
\label{figS1}
			
		\end{center}
	\end{figure}

\begin{figure*}[t]
		\begin{center}
			
			\includegraphics[width=180mm]{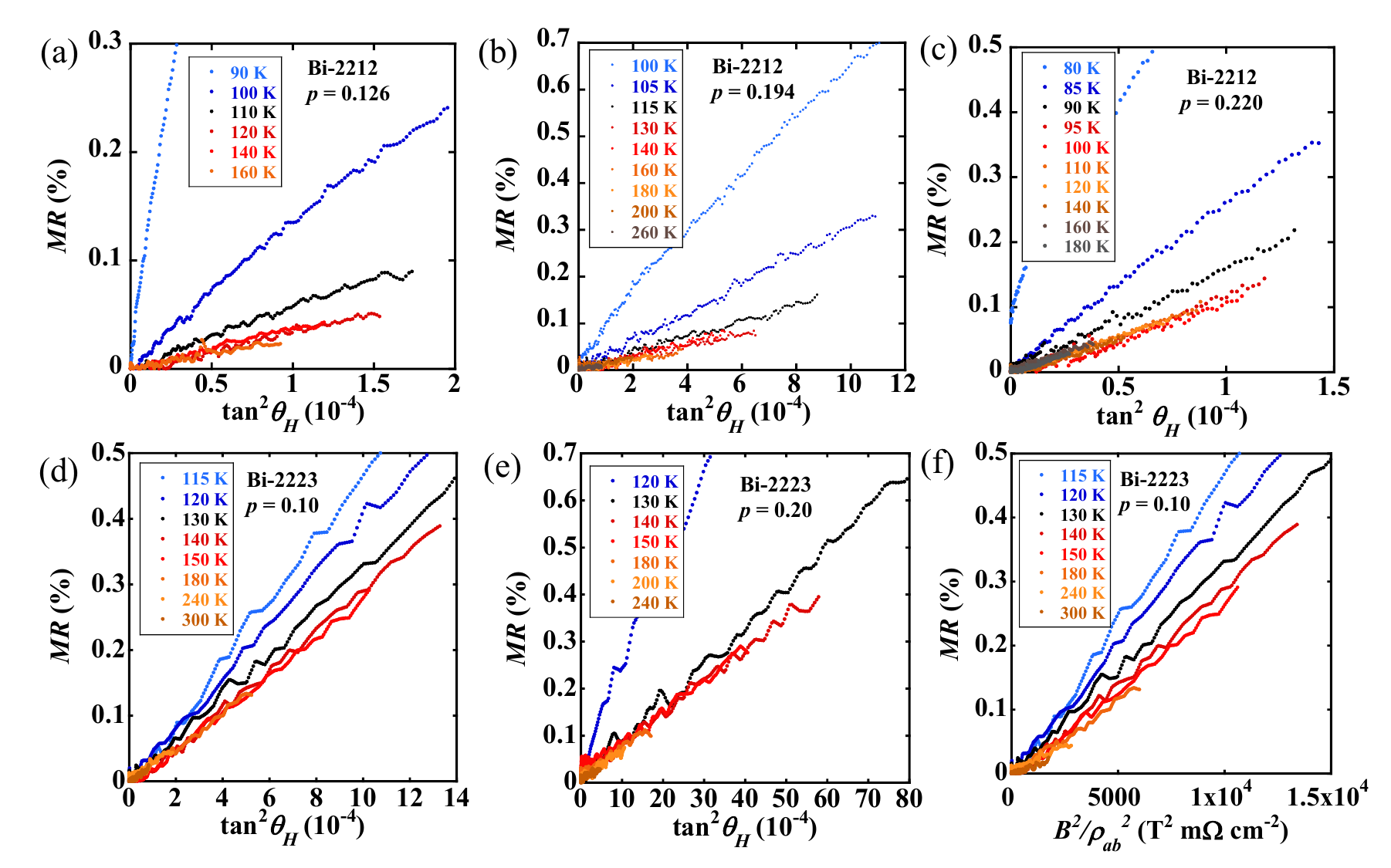}
			\caption{(Color online)
Magnetoresistance $MR$ versus $\tan^2\theta_H$ for (a) Bi-2212 ($p$ = 0.126), (b) Bi-2212 ($p$ = 0.194), (c) Bi-2212 ($p$ = 0.220), (d) Bi-2223 ($p$ = 0.10), and (e) Bi-2223 ($p$ = 0.20). 
(f) $MR$ versus $B^2/\rho_{ab}^2$ for Bi-2223 ($p$ = 0.10). In all panels, the temperatures corresponding to the black dots indicate $T_{scf}$.
}
\label{figS1}
			
		\end{center}
	\end{figure*}

\begin{figure*}[t]
		\begin{center}
			
			\includegraphics[width=180mm]{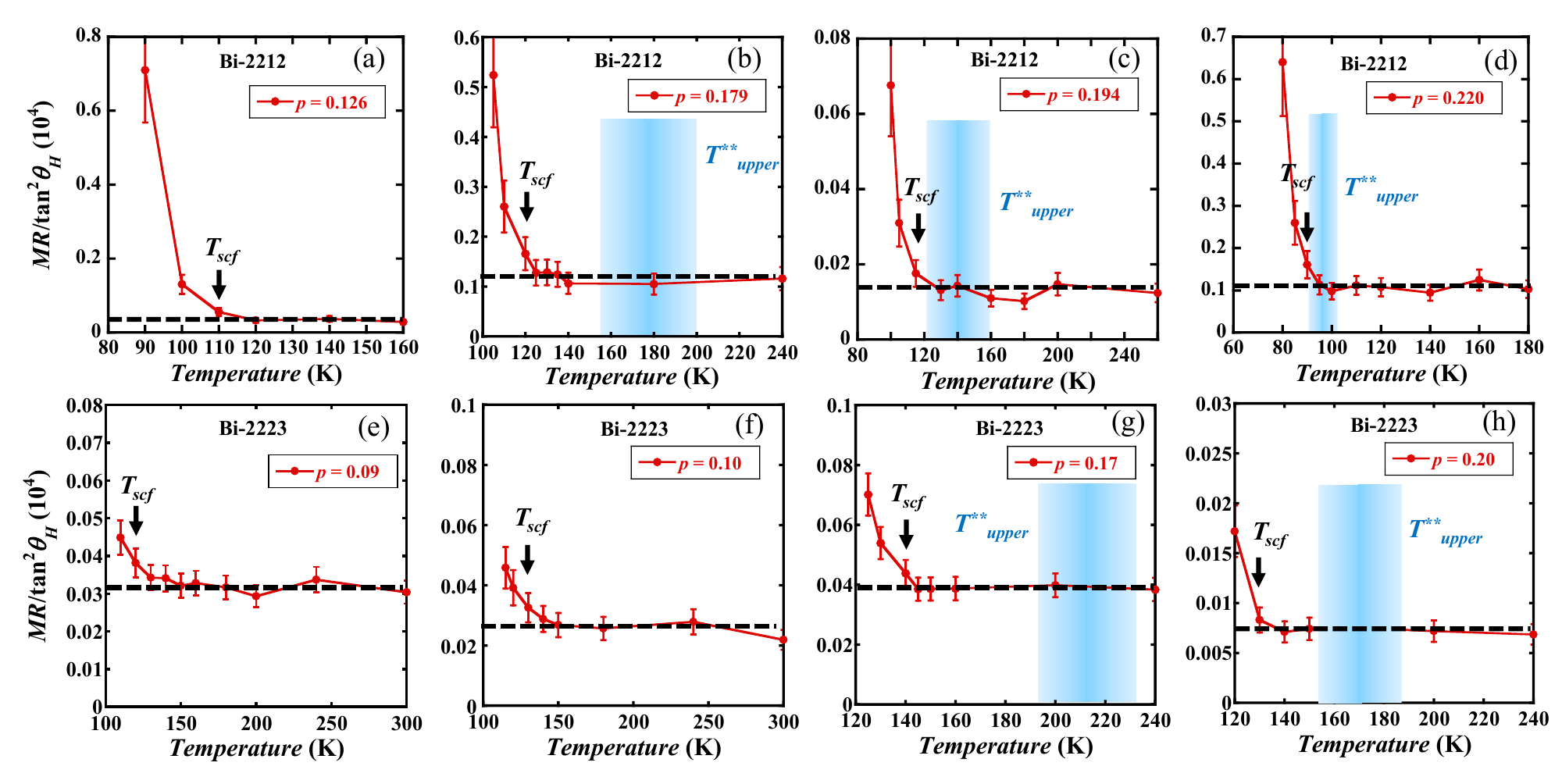}
			\caption{(Color online)
$MR$/$\tan^2\theta_H$ as a function of temperature for Bi-2212 with $p$ = (a) 0.126, (b) 0.179, (c) 0.194, and (d) 0.220, and for Bi-2223 with $p$ = (e) 0.09, (f) 0.10, (g) 0.17, and (h) 0.20. Arrows mark the onset temperatures of superconducting fluctuations, $T_{\mathrm{scf}}$. The blue shaded region indicates the uncertainty in $T^{**}_{\mathrm{upper}}$.
}
\label{figS1}
			
		\end{center}
	\end{figure*}

\begin{figure*}[t]
		\begin{center}
			
			\includegraphics[width=180mm]{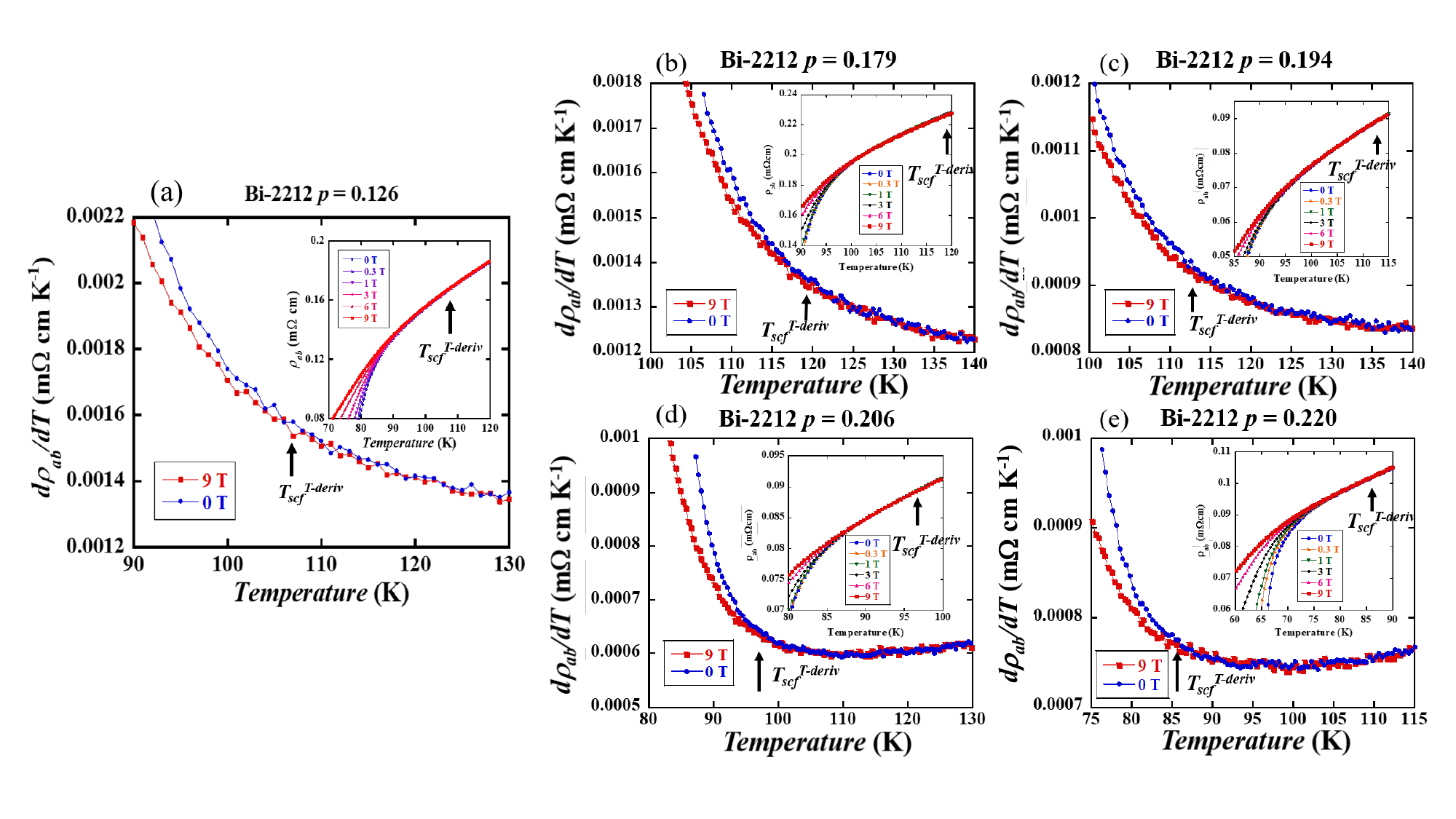}
			\caption{(Color online)
$d\rho_{ab}/dT$ as a function of temperature at 0 T and 9 T for Bi-2212 with $p$ = (a) 0.126, (a) 0.179, (b) 0.194, (c) 0.206, and (d) 0.220. Arrows mark the superconducting fluctuation onset temperatures $T_{\mathrm{scf}}^{\;\mathrm{T-deriv}}$, defined as the point where the 9 T curve decreases by more than 1$\%$ relative to the 0 T curve. Insets show expanded plots of $\rho_{ab}(T)$ near $T_c$ under magnetic fields up to 9 T. Panels (b)–(e) are reproduced from Ref. \cite{Watanabe22}.
}
\label{figS1}

		\end{center}
	\end{figure*}

\begin{figure*}[t]
		\begin{center}
			
			\includegraphics[width=150mm]{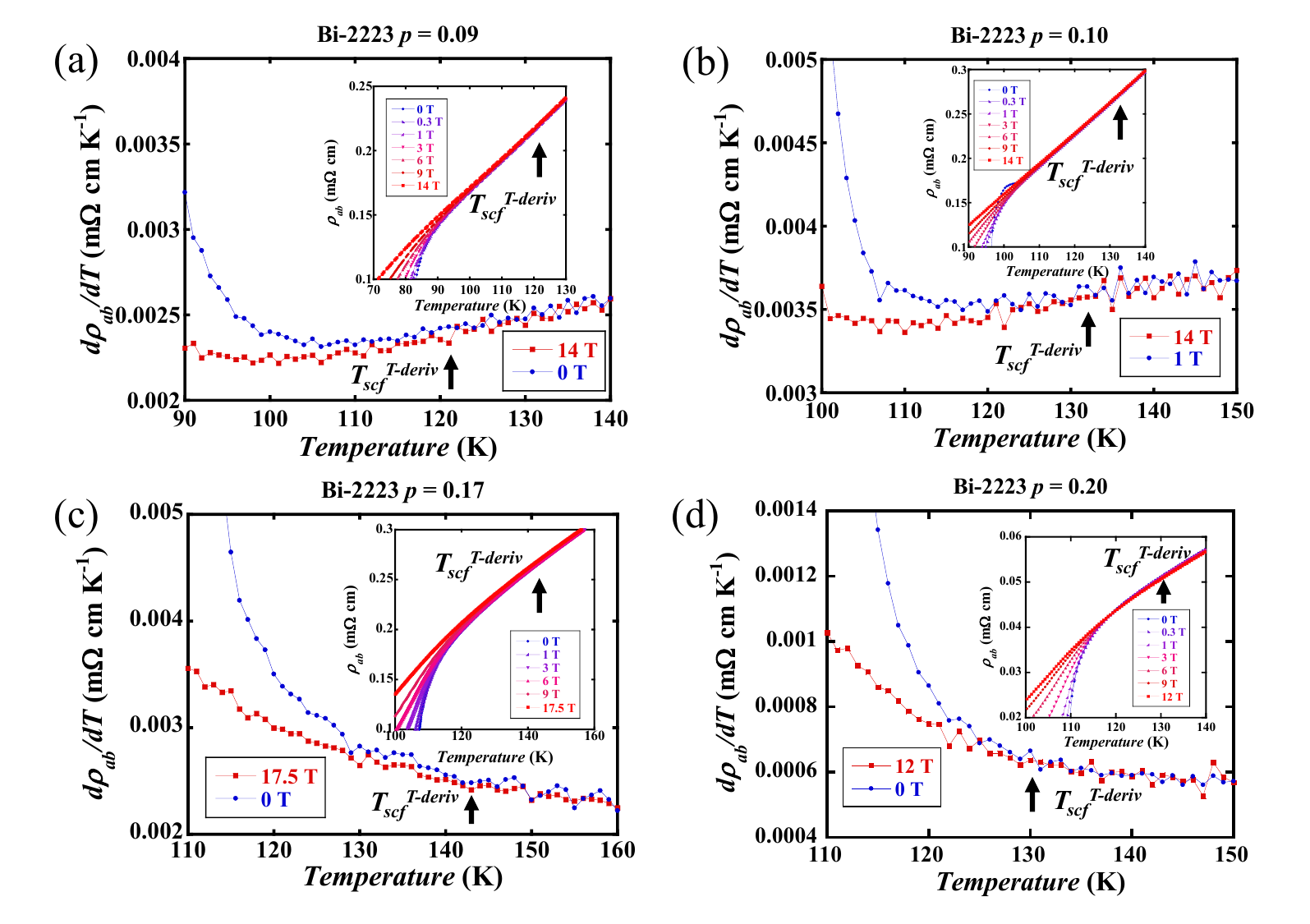}
			\caption{(Color online)
$d\rho_{ab}/dT$ as a function of temperature at 0 T and at the highest applied magnetic fields for Bi-2223 with $p$ = (a) 0.09, (b) 0.10, (c) 0.17, and (d) 0.20. Arrows mark the superconducting-fluctuation onset temperatures $T_{\mathrm{scf}}^{\;\mathrm{T-deriv}}$, defined as the point where the high field curve decreases by more than 1$\%$ relative to the 0 T curve. Insets show expanded plots of $\rho_{ab}(T)$ near $T_c$ under magnetic fields up to the highest applied values.
}
\label{figS1}
			
		\end{center}
	\end{figure*}

\section*{B. Procedures used to identify the pseudogap temperatures}

In clean cuprate systems with structurally flat CuO$_2$ planes—such as Hg 1201 \cite{Barisic13}, YBCO \cite{Ito93}, and multilayer Hg-1223 \cite{Carrington94}—the in-plane resistivity exhibits a robust $T$-linear behavior at high temperatures above the pseudogap temperature $T^*$. This $T$-linear regime is widely recognized as a hallmark of the strange-metal state and provides a reliable baseline for identifying deviations associated with pseudogap formation.

Underdoped Bi-2212 ($p$ = 0.126) and near optimally doped Bi-2223 ($p$ = 0.17) samples show the same characteristic high-temperature $T$-linear behavior. Therefore, following the criterion established by Ref. \cite{Ito93} for YBCO, the deviation from the high-temperature linear resistivity is an appropriate and physically justified method for determining $T^{**}_{\mathrm{upper}}$ in these systems.

The pseudogap opening temperature $T^{**}_{\mathrm{upper}}$ for these samples was determined using a 1$\%$ deviation from high-temperature $T$-linear behavior. The obtained $T^{**}_{\mathrm{upper}}$ values are 206 K and 213 K for Bi-2212 ($p$ = 0.126) and Bi-2223 ($p$ = 0.17), respectively.
Figure 1 presents a sensitivity analysis showing how the determination of $T^{**}_{\mathrm{upper}}$ changes when the deviation threshold is varied among 0.5$\%$, 1$\%$, and 2$\%$ for (a) Bi-2212 ($p$ = 0.126) and (b) Bi-2223 ($p$ = 0.17).
For Bi-2212, the resulting $T^{**}_{\mathrm{upper}}$ values are 225 K, 206 K, and 183 K; for Bi‑2223, they are 229 K, 213 K, and 186 K. Based on this analysis, we assign error bars that encompass the full range obtained from threshold values between 0.5$\%$ and 2$\%$.

Figure 2(a) shows $d\rho_{ab}/dT$ as a function of temperature for nearly optimally doped Bi-2212 ($p$ = 0.179) and overdoped Bi-2212 ($p$ = 0.194, 0.206, and 0.220). 
For these samples, $T^{**}_{\mathrm{upper}}$ was defined as the temperature at which $d\rho_{ab}/dT$ reaches its minimum. 
The resulting values of $T^{**}_{\mathrm{upper}}$ are 180, 140, 108, and 95 K for Bi-2212 with $p$ = 0.179, 0.194, 0.206, and 0.220, respectively.

Figure 2(b) shows $d\rho_{ab}/dT$ as a function of temperature for overdoped Bi-2223 ($p$ = 0.20) and underdoped Bi-2223 ($p$ = 0.09 and 0.10). 
For $p$ = 0.20, $T^{**}_{\mathrm{upper}}$ was defined in the same way as overdoped Bi-2212, yielding 170 K.
For underdoped Bi-2223, the characteristic pseudogap temperature $T^{**}_{\mathrm{lower}}$ was defined as the temperature at which $d\rho_{ab}/dT$ reaches its maximum, giving  173 K and 161 K for $p$ = 0.09 and 0.10, respectively.
In the case of $p$ = 0.09, a minimum appears at approximately 110 K, which lies below $T^{**}_{\mathrm{lower}}$ and corresponds to $T_{scf}$. 

As noted in Ref. \cite{Ando041}, determining $T^{**}_{\mathrm{upper}}$ and $T^{**}_{\mathrm{lower}}$ from the minima and maxima of the derivative is more objective than using deviations from linearity. However, the extrema in our data are broad, and noise introduces uncertainty in locating the precise temperature of the minimum or maximum. Therefore, we assign error bars that cover the temperature range over which the extrema remain constant within the noise level.

The criterion based on the minimum of the out-of-plane resistivity $\rho_{c}(T)$ provides an independent and well-established method for identifying the pseudogap onset. Figure 3 shows $\rho_{c}(T)$ for Bi-2212 ($p$ = 0.179, 0.194, 0.206, and 0.220) and Bi-2223 ($p$ = 0.20). Because the opening of the pseudogap causes $\rho_{c}$ to increase rapidly, the temperature at which $\rho_{c}$ reaches a minimum is taken as $T_{\mathrm{upper}}^{**\quad\rho_{c}}$, following Ref. \cite{Watanabe22}. The resulting values are 189, 138, 106, 91 K for Bi-2212 and 194 K for Bi-2223.

Figure 4 compares $T^{**}_{\mathrm{upper}}$ obtained from in-plane transport with $T_{\mathrm{upper}}^{**\quad\rho_{c}}$ obtained from $\rho_{c}(T)$. The two determinations coincide across all doping levels, confirming that the pseudogap onset extracted from the minimum in $d\rho_{ab}/dT$ is supported by an independent transport observable.

Regarding samples where the minimum in $d\rho_{ab}/dT$ appears broad (e.g., Bi-2212 with $p$ = 0.179 and 0.194, and Bi-2223 with $p$ = 0.20), the corresponding $\rho_{c}(T)$ curves exhibit clear minima, and the temperatures at which $d\rho_{ab}/dT$ becomes flat within experimental uncertainty fall within the same range. These intervals are shown as horizontal bars in Fig. 3 and lie well within the error bars used in the phase diagrams (see also Figs. 2 and 4). Thus, even when the minimum in $d\rho_{ab}/dT$ is broad, the determination of $T^{**}_{\mathrm{upper}}$ remains objective and reproducible.

Finally, we have compared our results with previously reported scanning tunneling spectroscopy (STS) measurements \cite{Oda02}. STS studies report pseudogap opening temperatures $T^*$ of approximately 200 K, 180 K, and 120 K for Bi-2212 at $p$ $\approx$ 0.12, 0.14, and 0.21, respectively. Our transport-derived values—206 K, 180 K, and 108 K for $p$ = 0.126, 0.179, and 0.206—are in excellent agreement. ARPES studies report a similar $T^* (p)$ relationship \cite{Vishik18}. These comparisons demonstrate that the pseudogap temperatures obtained from transport measurements are consistent with those derived from spectroscopic techniques.

Together, these results demonstrate that our method for identifying the pseudogap onset is robust and corroborated by independent measurements.

\section*{C. Temperature dependence of the in-plane resistivity for underdoped Bi-2223}

Figure 5 shows the in-plane resistivity $\rho_{ab}(T)$ of underdoped Bi-2223 ($p$ = 0.09) plotted as a function of $T^2$. 
A clear $\rho_{ab} \propto T^2$ dependence is observed between 130 K and 140 K, well within the pseudogap state 
($T \le T^{**}_{\mathrm{lower}}$). 
This behavior is consistent with that reported for Hg‑1201 \cite{Barisic13}, although the temperature range exhibiting 
$T^2$ dependence is narrower in Bi-2223. 
Since both Hg‑1201 and Bi-2223 are clean systems with structurally flat CuO$_2$ planes, the $T^2$ dependence of $\rho_{ab}(T)$ may represent an intrinsic property of underdoped cuprates.
%those reported for clean materials with flat CuO$_2$ planes such asand YBCO \cite{Ito93}

\section*{D. Additional magnetoresistance data not shown in the main text }

Figures 6(a)–6(e) show the magnetoresistance $MR$ for Bi-2212 ($p$ = 0.126, 0.194, and 0.220) and Bi-2223 ($p$ = 0.10 and 0.20) plotted as a function of $\tan^2\theta_H$. 
For each sample, the data collapse onto a single curve above 120, 130, 95, 140, and 140 K, respectively, indicating that the modified Kohler’s rule [Eq. (4) in the main text] is satisfied above these temperatures. 
Below these temperatures, an additional contribution to $MR$, attributable to superconducting fluctuations (SCF), becomes evident. 
From this behavior, the onset temperature of SCF, $T_{\mathrm{scf}}$, is estimated to be 110, 115, 90, 130, and 130 K for Bi-2212 ($p$ = 0.126, 0.194, and 0.220) and Bi-2223 ($p$ = 0.10 and 0.20), respectively.

Figure 6(f) shows a Kohler plot of the same data as in Fig. 6(d). 
The degree of overlap above 140 K is noticeably worse than in Fig. 6(d), demonstrating that the conventional Kohler’s rule is violated and that the modified Kohler’s rule provides a superior scaling description.

\section*{E. Procedures used to identify the onset temperature of superconducting fluctuations}

Figure 7 shows the slopes of $MR$ with respect to $\tan^2\theta_H$ as a function of temperature for (a)–(d) Bi-2212 ($p$ = 0.126, 0.179, 0.194, and 0.220) and (e)–(h) Bi-2223 ($p$ = 0.09, 0.10, 0.17, and 0.20). At high temperatures, the slopes remain nearly constant, demonstrating that the modified Kohler’s rule is satisfied. Upon cooling toward  $T_c$, the slopes begin to increase rapidly, reflecting the emergence of SCF effects.

To determine $T_{\mathrm{scf}}$ quantitatively, we define it as the temperature at which the slope exceeds the high-temperature constant value by more than 10$\%$. This criterion provides an objective and reproducible measure of the deviation point. In addition, the pseudogap temperatures $T^{**}_{\mathrm{upper}}$ are shown in the same panels. Importantly, the modified Kohler’s rule remains valid across $T^{**}_{\mathrm{upper}}$, reinforcing our conclusion that the pseudogap is distinct from SCF.

To corroborate our method for determining the onset of SCF, we examine the temperature dependence of the in-plane resistivity. Because SCF are readily suppressed by magnetic fields, comparing the resistivity with and without magnetic fields provides a sensitive and independent means of detecting their onset \cite{Watanabe22}. Figures 8(a)–8(e) show $d\rho_{ab}/dT$ at 0 T and 9 T for Bi-2212 ($p$ = 0.126, 0.179, 0.194, 0.206, and 0.220). Since 9 T most effectively suppresses SCF, we define $T_{\mathrm{scf}}$ as the temperature at which $d\rho_{ab}/dT$ at 9 T decreases by more than 1$\%$ relative to the zero field value. The resulting values are 107, 119, 113, 97, and 86 K for these doping levels. A similar analysis yields $T_{\mathrm{scf}}$ = 121, 132, 143, and 130 K for Bi 2223 with $p$ = 0.09, 0.10, 0.17, and 0.20, respectively (Fig. 9). We refer to these values as $T_{\mathrm{scf}}^{\;\mathrm{T-deriv}}$.

Figures 4(a) and 4(b) compare $T_{\mathrm{scf}}^{\;\mathrm{T-deriv}}$ with the $T_{\mathrm{scf}}$ obtained from the modified Kohler analysis in Fig. 7. The two determinations coincide across all doping levels, demonstrating that the deviation from the modified Kohler’s rule indeed corresponds to the onset of SCF. 

One might argue that these deviations from the modified Kohler’s rule could arise from mechanisms other than SCF, such as changes in the antiferromagnetic correlation length $\xi_{AF}$ due to charge-density-wave (CDW) formation. However, within the current-vertex-corrections (CVC) framework, the magnetoresistance $MR$ does \textit{not} depend on $\xi_{AF}$ [Eq. (4) in the main text]. Therefore, changes in $\xi_{AF}$ associated with CDW formation cannot produce deviations from the modified Kohler’s rule in this theoretical framework. Furthermore, CDW order is absent in overdoped Bi 2212 \cite{Loret20} and Bi 2223, yet the SCF related deviation is clearly observed in this regime. In addition, CDW signatures and changes in $\xi_{AF}$ are not field-sensitive in the manner characteristic of SCF. These considerations rule out CDW-related mechanisms as the origin of the observed deviation.

We note that the deviation of $MR$ from the high temperature modified Kohler curve (Fig. 3 in the main text and Fig. 6 in the Supplemental Material) always appears whenever SCF are present. If 14 T is not sufficient to suppress SCF, $MR$ continues to increase with magnetic field, and the slope $MR/\tan^2\theta_H$ remains larger than that of the high temperature modified Kohler curve. If 14 T were sufficient to suppress SCF, $MR$ would initially increase above the modified Kohler curve but would eventually saturate to the same slope once the field exceeds the upper critical field $B_{c2}$.
All of our experimental data exhibit the former behavior below $T_{\mathrm{scf}}$, indicating that—even at $T_{\mathrm{scf}}$—14 T is not sufficient to fully suppress SCF. Access to higher magnetic fields would allow a direct determination of $B_{c2}(T)$. Thus, the modified Kohler analysis is not only a sensitive and reliable method for detecting the onset of SCF, but also provides an opportunity to explore the underlying physics governing the field suppression of SCF.